\documentclass[aps, prd, twocolumn, amsmath, floats, floatfix, superscriptaddress, nofootinbib]{revtex4-2}
\usepackage{graphicx}
\usepackage{epsfig}
\usepackage{amsmath,amsfonts,amssymb}
\usepackage{caption}
\usepackage{subcaption}
\usepackage{wrapfig}

\usepackage{verbatim}
\usepackage{float}
\usepackage{morefloats}
\usepackage[dvipsnames]{xcolor}
\usepackage{booktabs}
\usepackage[normalem]{ulem}
\usepackage{multirow}
\usepackage{makecell}
\usepackage{tabulary}
\usepackage{cancel}
\usepackage{xcolor}

\usepackage{blindtext}      
\usepackage{hyperref}       
\hypersetup{
	colorlinks   = true, 
	urlcolor     = blue, 
	linkcolor    = blue, 
	citecolor    = red,  
}
\usepackage{url}

\usepackage{booktabs}       
\usepackage{multirow}       
\usepackage{adjustbox}      
\usepackage{cellspace}
\usepackage{tablefootnote}  
\usepackage{physics}                           
\usepackage{enumitem}
\setlist[itemize]{noitemsep, topsep=0cm, leftmargin = 1.2cm}        
\setlist[enumerate]{noitemsep, topsep=0cm, leftmargin = 1.2cm}      

\usepackage{ragged2e} 
\usepackage{subcaption} 
\DeclareCaptionJustification{justified}{\justifying} 

\begin{document}

\title{Automatic classification pipeline for glitches in the Virgo detector}

\author{Tiago~Fernandes}
\thanks{Corresponding author}
\email{tiasafer@alumni.uv.es}
\affiliation{Centro de F\'{\i}sica das Universidades do Minho e do Porto (CF-UM-UP), Universidade do Minho, 4710-057 Braga, Portugal}
\affiliation{Departamento de
  Astronom\'{\i}a y Astrof\'{\i}sica, Universitat de Val\`encia,
  Avinguda Vicent Andrés Estellés 19, 46100, Burjassot (Val\`encia), Spain}

\author{Francesco Di Renzo}
\affiliation{Universit\`a di Firenze, I-50019 Sesto Fiorentino, Firenze, Italy}
\affiliation{INFN, Sezione di Firenze, I-50019 Sesto Fiorentino, Firenze, Italy}

\author{Antonio~Onofre}
\affiliation{Centro de F\'{\i}sica das Universidades do Minho e do Porto (CF-UM-UP), Universidade do Minho, 4710-057 Braga, Portugal}

\author{Alejandro \surname{Torres-Forn\'e}}
\affiliation{Departamento de
  Astronom\'{\i}a y Astrof\'{\i}sica, Universitat de Val\`encia,
  Avinguda Vicent Andrés Estellés 19, 46100, Burjassot (Val\`encia), Spain}
\affiliation{Observatorio Astron\'omico, Universitat de València, C/ Catedrático Jos\'e Beltr\'an 2, 46980, Paterna (València), Spain}

\author{Jos\'e A.~Font}
\affiliation{Departamento de
  Astronom\'{\i}a y Astrof\'{\i}sica, Universitat de Val\`encia,
  Avinguda Vicent Andrés Estellés 19, 46100, Burjassot (Val\`encia), Spain}
\affiliation{Observatorio Astron\'omico, Universitat de València, C/ Catedrático Jos\'e Beltr\'an 2, 46980, Paterna (València), Spain}

\begin{abstract}
Glitches frequently contaminate data in gravitational-wave detectors, complicating the observation and analysis of astrophysical signals. This work introduces \texttt{VIGILant}, an automatic pipeline for classification and visualization of glitches in the Virgo detector. Using a curated dataset of Virgo O3b glitches, two machine learning approaches are evaluated: tree-based models (Decision Tree, Random Forest and XGBoost) using structured \texttt{Omicron} parameters, and Convolutional Neural Networks (ResNet) trained on spectrogram images.
While tree-based models offer higher interpretability and fast training, the ResNet34 model achieved superior performance, reaching a F1 score of 0.9772 and accuracy of 0.9833 in the testing set, with inference times of tens of milliseconds per glitch. The pipeline has been deployed for daily operation at the Virgo site since observing run O4c, providing the Virgo collaboration with an interactive dashboard to monitor glitch populations and detector behavior. This allows to identify low-confidence predictions, highlighting glitches requiring further attention. 
\end{abstract}

\maketitle

\section{Introduction}

In the last ten years, since the first detection of a gravitational wave (GW) in 2015~\cite{LIGOScientific:2016aoc}, the LIGO-Virgo-KAGRA (LVK) collaboration~\cite{LIGOScientific:2014pky, VIRGO:2014yos, KAGRA} has announced 218 GW confident detections, all involving compact binary coalescences~\cite{TheLIGOScientificCollaboration2025}. These observations are significantly impacting relativistic astrophysics on many fronts. 
They enable testing of General Relativity in the strong-field regime of gravity~\cite{TheLIGOScientificCollaboration2026}, constrain the equation of state of neutron stars~\cite{Abbott2018}, reveal the population properties of compact binaries~\cite{The_LIGO_Scientific_Collaboration2025-qs}, and even provide new estimates of the Hubble constant~\cite{The_LIGO_Scientific_Collaboration2025-au}. 

The observation of GWs is intrinsically demanding due to the low signal-to-noise ratio (SNR) conditions. Making things more challenging, GW detectors are susceptible to transient non-Gaussian noises, which are much more frequent than astrophysical GW signals and potentially hinder the analyses.
These transient noises, dubbed ``glitches", have environmental, anthropogenic or instrumental origins, which are often unknown, and contaminate the data from the GW detectors~\cite{detectornoise}. Glitches can mimic real astrophysical signals, especially those produced in high-mass binary black hole mergers, or overlap with true astrophysical signals, which hinders both detection and parameter estimation of the source properties~\cite{Davis2021}. The glitch rate on the Virgo detector, for glitches with SNR $> 6.5$ and peak frequency between 20~Hz and 2048~Hz, was around 0.1 per minute during the second part of the O4 observation period~\cite{logbook_glitchrate}
and 29\% of the new GW events observed in the O4a period were found to be contaminated by glitches~\cite{TheLIGOScientificCollaboration2025}. Therefore, monitoring glitches and mitigating their harmful effect is a crucial activity for characterizing the detectors and increasing their sensitivity. 

The efforts to diagnose and mitigate glitches are the responsibility of the Detector Characterization groups in the LVK collaboration~\cite{Davis2021, Davis2022}, with Machine Learning (ML) approaches playing an increasingly important role in these studies~\cite{Cuoco:2024cdk}.
A key investigation involves the classification of glitches into classes, each of which has a distinctive morphology in its time-frequency representation. Grouping glitches into classes is helpful for noise hunting, as one can look for correlations between a class of glitches and the detectors’ auxiliary channels, which measure the status of the detector and the environment. The LVK collaboration uses the \texttt{Omicron} algorithm~\cite{Robinet2020} to identify and characterize transient sources of noise in the detectors. 
This algorithm produces multiple time-frequency planes of the data using Q transforms at different values and stacks them.
Using this representation, tiles with excess power, quantified by the SNR, are identified and characterized as unclustered triggers. These are then grouped based on temporal proximity to form clustered triggers or simply `triggers'. Each trigger is assigned the GPS time of its highest-power tile and further described by \texttt{Omicron} parameters such as peak frequency and SNR. Together, these quantities form the basis for subsequent glitch investigations. 
In the context of this work, and the classification frameworks detailed below, we focus on \texttt{Omicron} triggers generated from the main strain channel, rather than the auxiliary channels.

Another important tool for the LVK analysis is the Gravity Spy project and the associated algorithm~\cite{Zevin2017a}, 
which classifies the glitches observed in the strain data channels in a pre-defined set of glitch families which were reviewed by LIGO experts and found to be more frequent, as well as more feature characteristic, in LIGO data. 
For this, the method uses the GPS times from \texttt{Omicron} triggers with SNR above 7.5 to produce Q-transform representations centered around that time, with four different time durations (0.5, 1.0, 2.0 and 4.0 seconds). 
These are then fed to a previously trained ML algorithm, which assigns a glitch family to each glitch. Each label is also associated with a certain ``confidence'', which corresponds to the maximum value of the class probabilities. These glitch classifications, along with the respective \texttt{Omicron} trigger parameters, are saved in an internal database, the LigoDV-Web GlitchDB\footnote{\url{https://ldvw.ligo.caltech.edu/ldvw/gspySearch}}, and often uploaded for public access\footnote{For instance, in \url{https://zenodo.org/records/5649212}.}.

Despite all the advances facilitated by the Gravity Spy project, its ML classifier is known for being too optimistic in its confidence scores, and even making wrong predictions with very high confidence~\cite{Zevin2023, Wu2024}. This issue is even more problematic for Virgo glitches, as the training corpus of the ML model consists only of LIGO glitches, and the families are mostly LIGO-oriented. 
Given that the differences between Virgo and the LIGO detectors are more substantial than those between the two LIGO instruments, particularly with respect to hardware design and operational sensitivity, it is expected that the glitch population observed in Virgo differs from that observed in LIGO. 
Consequently, some classes of glitches observed in Virgo can overlap with ones in LIGO but others may constitute an out-of-distribution dataset for the Gravity Spy classifier, for which neither the class definitions nor the associated confidence estimates are expected to remain reliable.
Moreover, overconfident but incorrect classifications can mislead detector characterization studies, biasing the identification of noise sources and the interpretation of data-quality investigations for the Virgo detector.

The need for Virgo-focused glitch characterization has been previously recognized, for example in the GWitchHunters project~\cite{Razzano2023}, which explored citizen-science–based approaches to the study of Virgo glitches. However, these efforts did not result in a continuously deployed, automated classification pipeline integrated into Virgo detector characterization workflows.
This is the aim of this work. Here, we present a Machine Learning–based pipeline dubbed \texttt{VIGILant} (\textbf{Vi}rgo \textbf{G}litch \textbf{I}dentification and \textbf{L}earning) for the automated classification of Virgo glitches, with the resulting classifications organized in an interactive dashboard accessible within the collaboration. The current version of \texttt{VIGILant} has been operating daily since observing run O4c, monitoring glitch populations and detector behavior and identifying glitches requiring further attention. Future extension will complement the pipeline's capabilities through the incorporation of unsupervised clustering for novel glitch discovery and advanced reconstruction methods for noise mitigation.

This paper is organized as follows: in Section~\ref{sec: methods} we describe the construction of the glitch dataset used in this work and introduce the different ML algorithms employed. Section~\ref{sec: model training} discusses the training of the different models and the selection of the best one, while Section~\ref{sec: model eval} describes the evaluation of the selected model and studies its generalization ability. In Section~\ref{sec: pipeline} we describe the integration of the classification model into a pipeline deployed on the Virgo computing cluster. Finally, we present the main conclusions  of this work as well as future developments of the pipeline in Section~\ref{sec: conclusion}.

\section{Methods}
\label{sec: methods}

\subsection{Dataset}

For this work, Virgo detector glitches from the second half of the third observing run (O3b) were used. These glitches were sourced from the GlitchDB, filtering for the O3b observing run and a GPS time from the beginning (2019-11-01 15:00:00 UTC) to the end (2020-03-27 17:00:00 UTC) of O3b.

\begin{figure}
    \centering
    \includegraphics[scale=0.39]{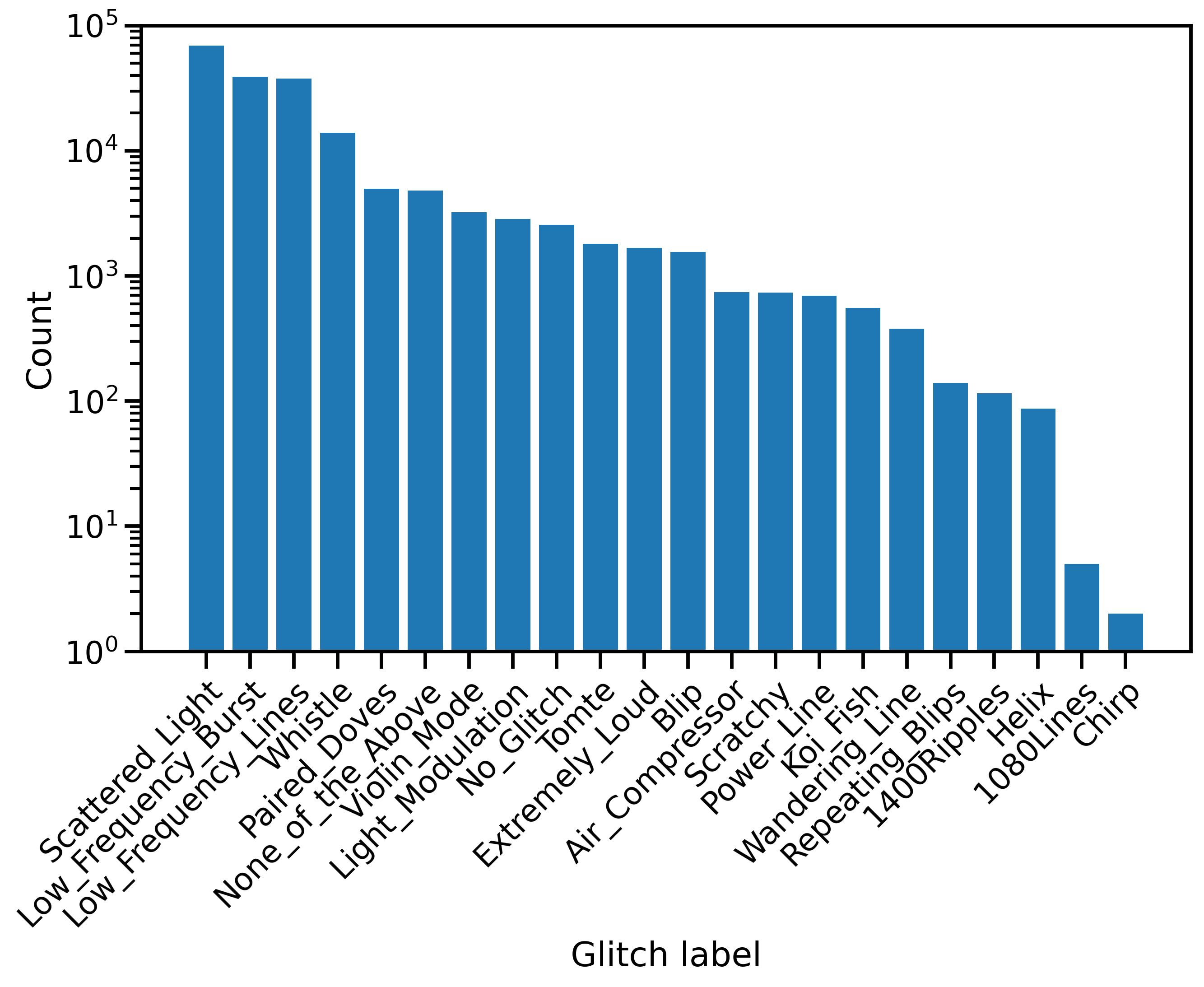}
    \caption{Distribution of the Gravity Spy labels for Virgo O3b glitches. There are 22 classes, with a highly imbalanced representation.}
    \label{fig: old_labels}
\end{figure}

We started from the Gravity Spy labels, whose distribution in our data is shown in Figure~\ref{fig: old_labels}. We can observe that there are 22 classes and that while some have less than 10 examples, the three main classes represent more than three quarters of the total of glitches. This high imbalance can be detrimental for the training of ML models, so this was taken into account when creating the final dataset and evaluating the performance of the models.

We visually inspected some spectrograms for each of the classes at different confidence levels to better understand the accuracy of the Gravity Spy labels. We found that no clear distinction between \texttt{Scattered\_Light}, \texttt{Low\_Frequency\_Burst} and \texttt{Low\_Frequency\_Lines} was possible, even without having to define a minimum confidence. This was also the case for the \texttt{Light\_Modulation} and \texttt{Paired\_Doves} classes if the minimum Gravity Spy confidence was set to 0.9. This is also supported by the fact that the vast majority of the samples from these classes is relabeled as \texttt{Scattered\_Light} in the dataset using the volunteer classifications to complement the ML model\footnote{Available at \url{https://zenodo.org/records/5911227}.}. 
In addition, we also observed that all samples of the \texttt{Whistle} class (the fourth most represented) showed a horizontal feature between 370 and 500 Hz. For frequencies closer to 500 Hz, this feature is consistent with the beating of an excited sideband with the carrier frequency of a violin mode from the Virgo mirror suspensions. At lower frequencies, however, the feature cannot be attributed to violin modes, although a similar mechanism associated with other spectral lines may be responsible.
For other classes like \texttt{No\_Glitch}, \texttt{Tomte}, \texttt{Extremely\_Loud}, \texttt{Blip} and \texttt{Koi\_Fish}, we found the labels to be mostly correct, especially as the confidence level was increased.

\begin{figure*}[!tbp]
    \centering
    \includegraphics[scale=0.48]{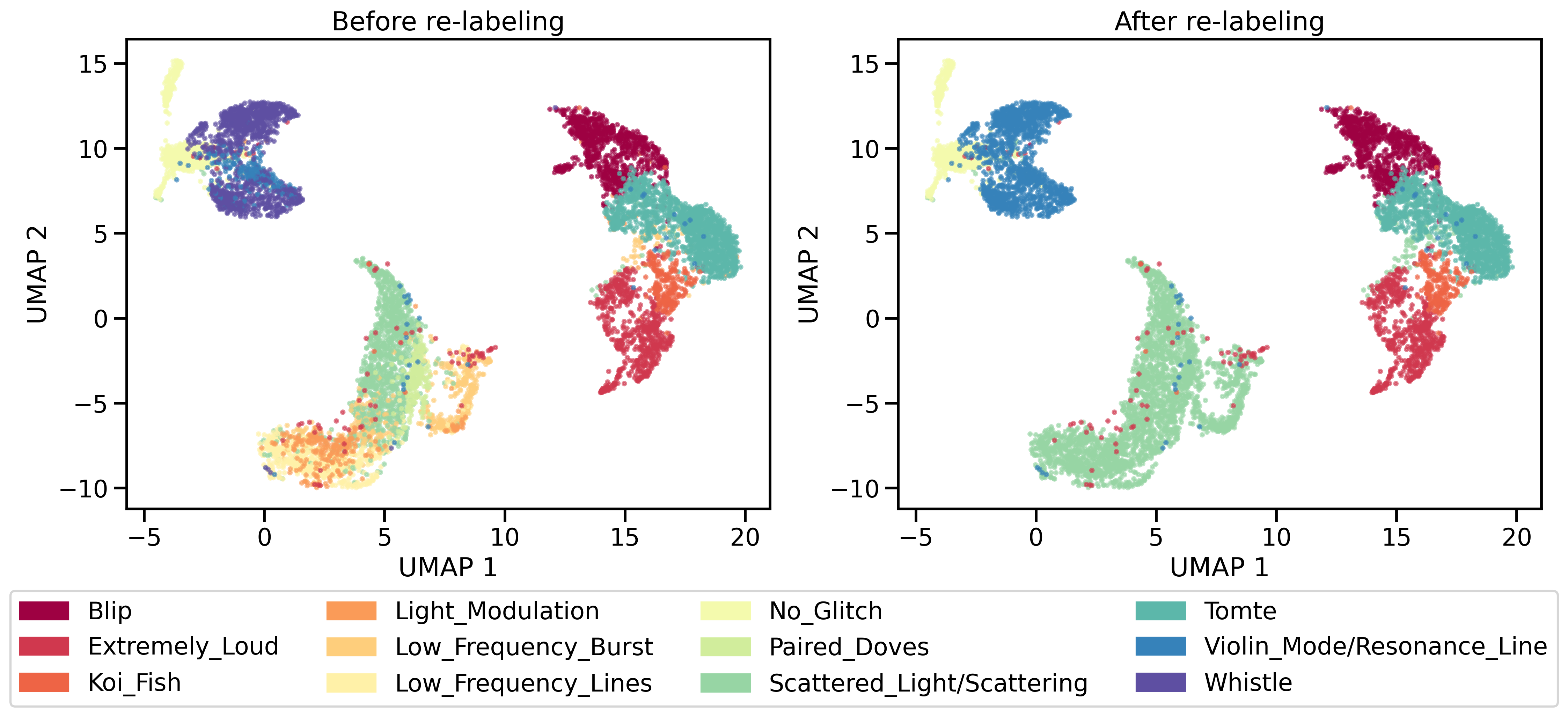}
    \caption{UMAP projection of the curated glitch dataset, before and after the re-labeling of some glitches. The re-labeling grouped two sets of closely related glitch families into two single families.}
    \label{fig: umap}
\end{figure*}

To better showcase the justification behind the re-labeling of some glitch classes, we projected the raw pixel values of the glitch spectrograms (which will be introduced in Section~\ref{subsection: spectrograms}) using the Uniform Manifold Approximation and Projection (UMAP) algorithm~\cite{McInnes2018-xo} to represent the highly dimensional glitches in two dimensions. 
Due to the number of pixels (3 channel dimensions $\times$ 140 $\times$ 170), we performed an intermediate step to reduce the feature dimension. Using Principal Component Analysis (PCA)~\cite{Jolliffe2002Principal} we reduced the data to 100 features, which kept 79.3\% of the total variance.
We show the UMAP projection of our training set in two dimensions in Figure~\ref{fig: umap}, with the original Gravity Spy labels (on the left), and with the new labels we chose (on the right).
There is a lot of intersection between the \texttt{Violin\_Mode} and \texttt{Whistle} samples in the plot on the left, which justifies their inclusion in a single class we named \texttt{Resonance\_Line}. 
Similarly, there is also a lot of superposition between the \texttt{Scattered\_Light}, \texttt{Paired\_Doves}, \texttt{Light\_Modulation}, \texttt{Low\_Frequency\_Burst} and \texttt{Low\_Frequency\_Lines}, which we joined in a single class named \texttt{Scattering}. 

Using these insights, we set to build a dataset with around 10,000 glitches, while avoiding having classes with too few glitches. 
Due to its catch-all nature, we also decided to drop the \texttt{None\_of\_the\_Above} class, which was first idealized to be able to classify glitches which do not belong to any of the other families, but was dropped in the most recent iterations of the Gravity Spy project~\cite{Wu2024}\footnote{This is because when training the ML classifier using a supervised training framework, the \texttt{None\_of\_the\_Above} class is treated in the same way as the others, i.e.~the algorithm tries to find the common patterns of the examples of this class in the training dataset, preventing its generalization to unseen examples. 
We also note that it is possible that this class includes new families of glitches, which we plan to address in future developments of the algorithm.
}
.

\begin{figure}
    \centering
    \includegraphics[scale=0.39]{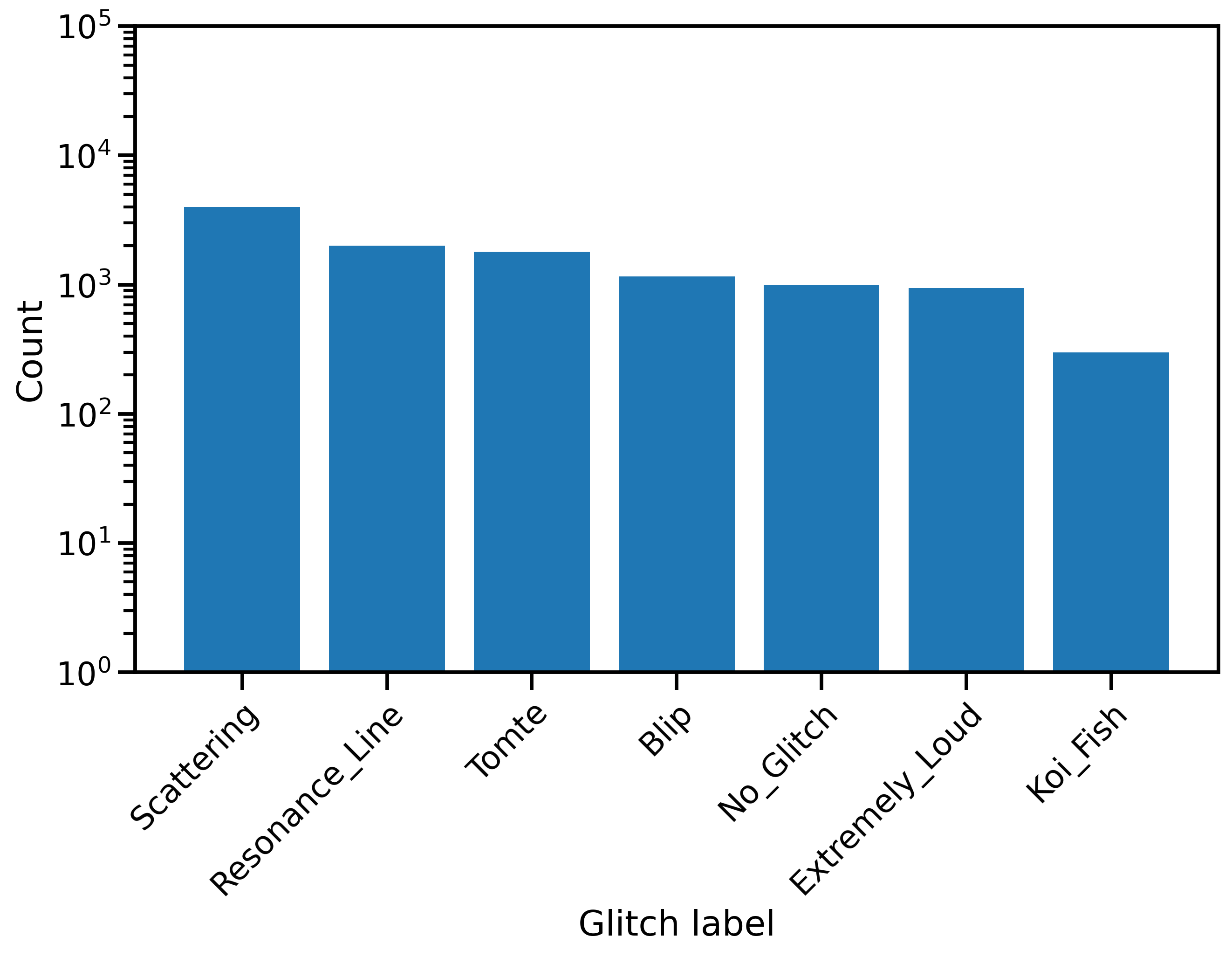}
    \caption{Distribution of the labels for our Virgo O3b glitch dataset. There are 7 classes, with less imbalance than in the original data.}
    \label{fig: new_labels}
\end{figure}

Finally, our Virgo O3b dataset has the following classes and constitution:
\begin{itemize}
    \item 4000 \texttt{Scattering} glitches, formed by the 309 \texttt{Paired\_Doves} with confidence $> 0.9$, the 360 \texttt{Light\_Modulation} glitches with confidence $> 0.9$ and 3331 glitches obtained using a weighted random sample (using confidence as the weight), from the \texttt{Scattered\_Light}, \texttt{Low\_Frequency\_Burst} and \texttt{Low\_Frequency\_Lines} classes;
    \item 2000 \texttt{Resonance\_Line} glitches obtained using weighted random sampling from the \texttt{Whistle} and \texttt{Violin\_Mode} classes;
    \item 1801 \texttt{Tomte} glitches, spanning all the glitches Gravity Spy classified with this label;
    \item 1156 \texttt{Blip} glitches, containing all examples classified by Gravity Spy with this label with a confidence of at least 0.9;
    \item 1000 \texttt{No\_Glitch} examples, corresponding to the most confident classifications of Gravity Spy for this family;
    \item 938 \texttt{Extremely\_Loud}, containing all glitches classified by Gravity Spy with this label with a confidence of at least 0.9;
    \item 298 \texttt{Koi\_Fish} glitches, spanning all the glitches Gravity Spy classified with this label with a confidence of at least 0.9.
\end{itemize}
In total, the dataset contains 11193 glitches distributed over 7~classes, as seen in Figure~\ref{fig: new_labels}, with the most represented class forming 35\% of the dataset and the minority class contributing to 2\% of the total. Although still imbalanced, the weight of the \texttt{Scattering} class is reduced in relation to the original data, and the minority class still contains almost 300 examples. The glitches were selected from the full O3b dataset without imposing temporal constraints. Since some glitch classes appear or disappear over periods of weeks, the specific examples included in the dataset depend on the time periods sampled. Therefore, we focus on the main classes with a sufficient number of examples to allow representative sampling.

The generated dataset was split in the usual training, validation and testing subsets for ML model training and selection, with a split ratio of 70\%, 15\% and 15\%, respectively. 
Stratified sampling was employed to preserve the class distribution in each subset, as seen in Table~\ref{tab: dataset_split}. 
The training set is used to train the ML models, while the validation dataset is used to evaluate their performance and guide  model selection. The test set was kept aside during the whole investigation, and was only used at the end to evaluate the selected model.

\begin{table}[t!]
    \setlength{\tabcolsep}{6pt}
    \centering
    \caption{Number of samples of each class in the three subsets of the Virgo O3b glitch dataset. Percentages (shown for the training set) indicate the fraction of each subset represented by the given class. Stratified sampling ensures that the class distribution is the same across all subsets.}
    \begin{tabular}{lccc}
    \hline
    \textbf{Class} & \textbf{Training} & \textbf{Validation} & \textbf{Testing}\\
    \hline
    \texttt{Scattering} & 2800 (35.7\%) & 600 & 600 \\
    \texttt{Resonance\_Line} & 1400 (17.9\%) & 300 & 300 \\
    \texttt{Tomte} & 1261 (16.1\%) & 270 & 270\\
    \texttt{Blip} & 810 (10.3\%) & 173 & 173 \\ 
    \texttt{No\_Glitch} & 700 (8.9\%) & 150 & 150 \\
    \texttt{Extremely\_Loud} & 656 (8.4\%) & 141 & 141 \\
    \texttt{Koi\_Fish} & 208 (2.7\%) & 45 & 45 \\
    \hline
    \textbf{Total} & 7835 & 1679 & 1679\\
    \hline
    \end{tabular}
    \label{tab: dataset_split}
\end{table}

\subsubsection{Omicron parameters}

For each glitch, identified by its central GPS time, a few parameters that are useful for its characterization are computed by the \texttt{Omicron} algorithm:
\begin{itemize}
    \item \texttt{peakFreq}: the peak frequency, that is, the frequency of the tile at which the SNR is maximized;
    \item \texttt{SNR}: the signal-to-noise-ratio of the loudest tile forming this glitch;
    \item \texttt{amplitude}: the amplitude estimation for the loudest tile;
    \item \texttt{centralFreq}: the central frequency, which is the average between the minimum and maximum frequencies of the glitch;
    \item \texttt{duration} the difference between the minimum and maximum times;
    \item \texttt{bandwidth}: the difference between the minimum and maximum frequencies of the glitch. 
\end{itemize}
More details on these parameters can be found in~\cite{Robinet2018}. 
These parameters alone provide a basic characterization of the glitches, which could be used for a rough classification using a simple algorithm. This motivated the experiments using the models presented in Section \ref{subsec: tree models}.

\subsubsection{Spectrograms}
\label{subsection: spectrograms}

\begin{figure}
    \centering
    \includegraphics[scale=0.45]{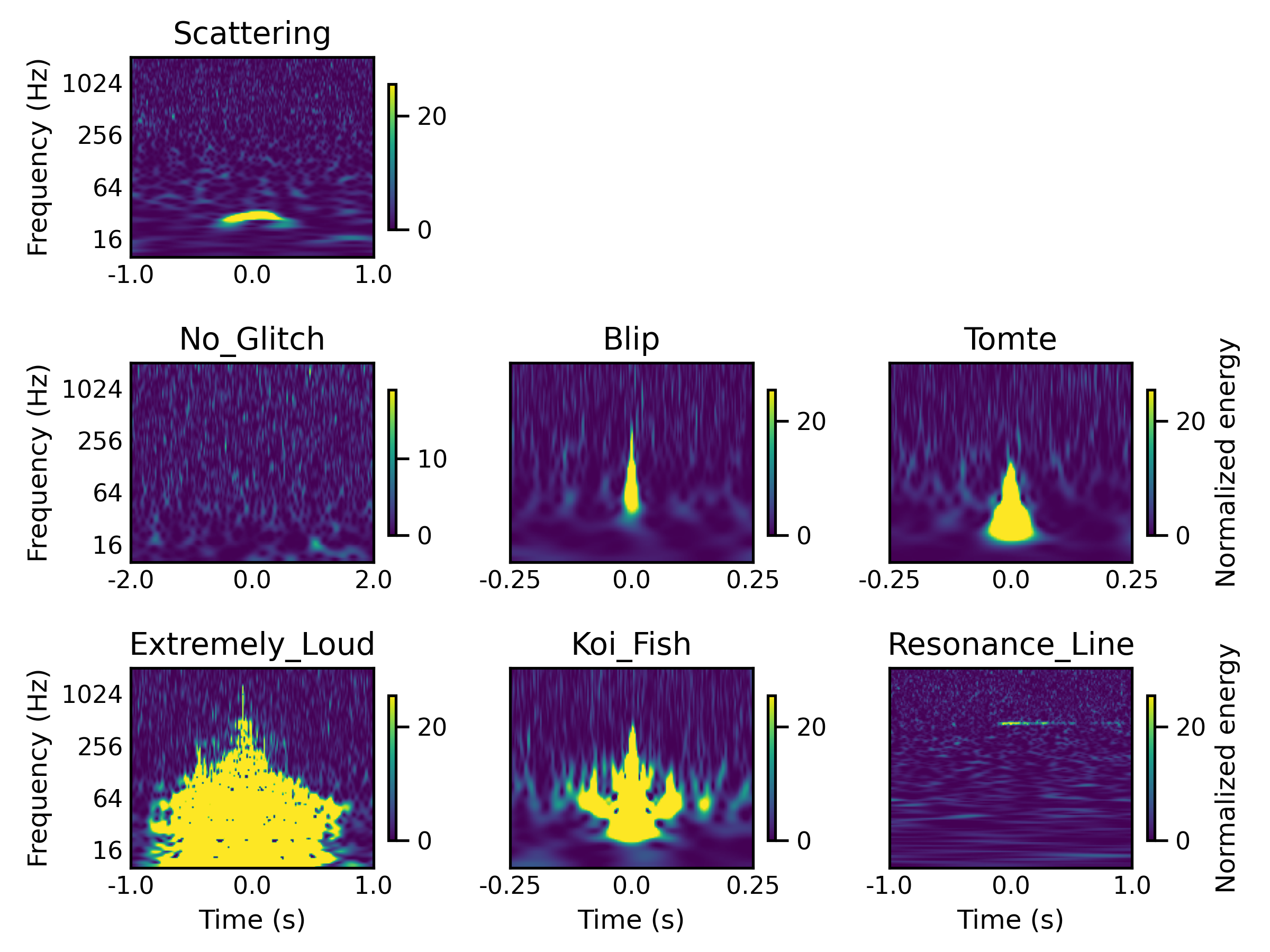}
    \caption{Examples of the different glitch classes in our Virgo O3b dataset.}
    \label{fig: glitch_examples}
\end{figure}

Aside from their \texttt{Omicron} parameters, glitches can be represented as spectrograms, using the Q-transform method~\cite{Chatterji2004}. In this representation, a high resolution spectrogram shows how the glitch frequency ($y$-axis) and amplitude (colour scale) evolve over time ($x$-axis). An example of each glitch class using this representation is show in Figure~\ref{fig: glitch_examples}. Spectrograms are generated using the 
\texttt{gwpy} library~\cite{gwpy}, with the default parameters. With this setup the Q-transform method tests four different Q values and chooses the Q for which the SNR of the loudest pixel is the highest.
Similarly to the Gravity Spy project, 64 seconds of data around the glitch GPS time are used for the whitening. However, the whitening is performed using a median average of overlapping periodograms, which is more robust to outliers than the Welch method (mean of periodograms) employed by Gravity Spy. We note, however, that for very loud glitches this can produce stronger saturation in whitened spectrograms, which may affect their visibility.
For each glitch, we generate three spectrograms, with three different time durations (0.5 s, 2.0 s and 4.0 s) centered on the glitch GPS time, because the different time scales allow to better identify different glitch morphologies. A maximum threshold for the normalized amplitude is set to 25.5. 

\begin{figure}
    \centering
    \includegraphics[scale=0.23]{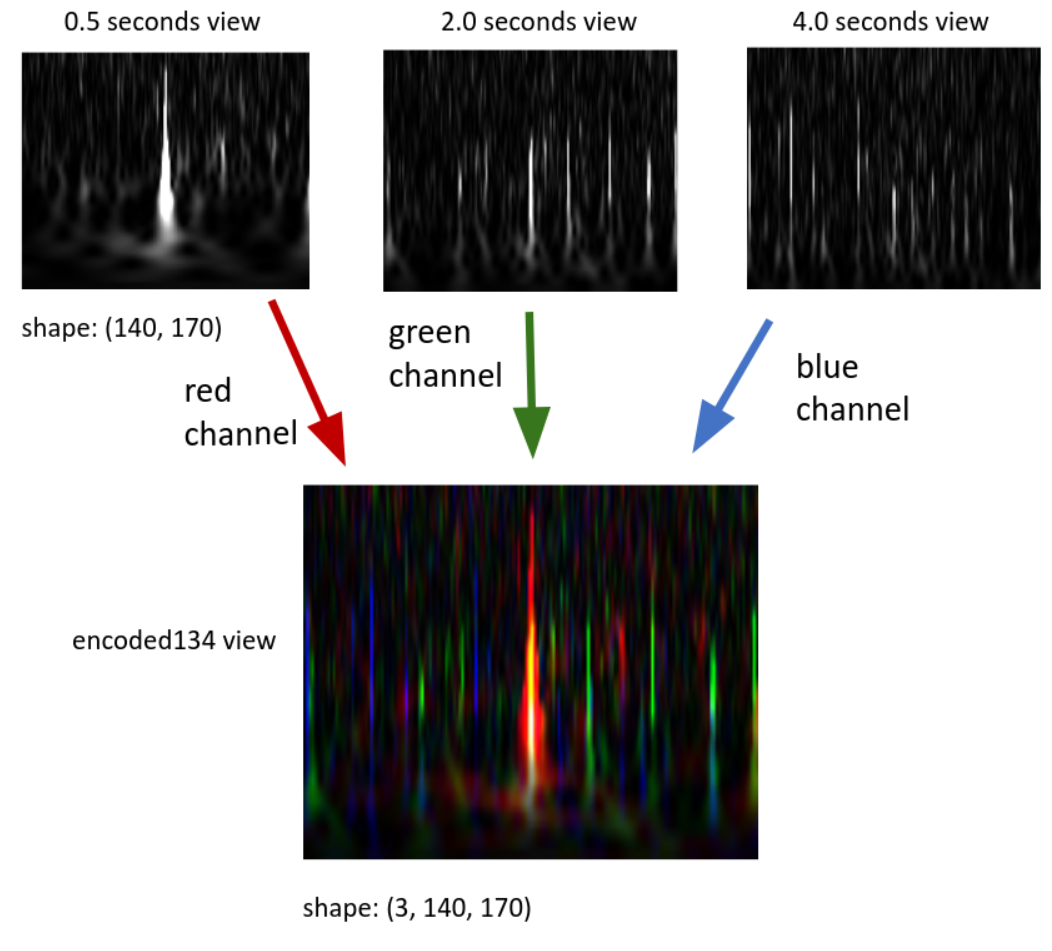}
    \caption{Scheme of the construction of the encoded134 view. The three grayscale spectrograms with different time windows are stacked along the colour dimension.}
    \label{fig: encoded134}
\end{figure}

These spectrograms are the input to the ResNet models presented in subsection~\ref{subsec: resnet models}. 
Given our previous findings in~\cite{Fernandes2023}, we save each glitch spectrogram as a $170 \times 140$ pixel grayscale image\footnote{Although we represent glitch images using a colour scale for better visualization, they are saved in grayscale, which only uses one colour channel and is thus more computationally efficient.}. 
Then, for each glitch, we use the image colour channel dimension to encode the three different time-duration spectrograms in a single image, forming the so-called ``encoded" views introduced in~\cite{George2017}. 
This encoded view, using the 0.5, 2.0 and 4.0 second windows, corresponds to the ``encoded134" view, which was found to be the best data representation in terms of accuracy and speed in~\cite{Fernandes2023}. Its creation is schematized in Figure~\ref{fig: encoded134}.

\subsection{Machine Learning models}

In this work we employ two groups of ML models for the glitch classification: (i) tree-based models trained on the \texttt{Omicron} features, and (ii) convolutional neural networks trained on the spectrogram images.

\subsubsection{Tree-based models}
\label{subsec: tree models}

Three tree-based ML algorithms were employed to classify glitches using the \texttt{Omicron} features: Decision Trees (DTs), Random Forests (RFs) and Extreme Gradient Boosting (XGBoost). These models were selected for their established performance on structured data and their interpretability.

The simplest algorithm, the Decision Tree~\cite{Breiman1984}, is trained by constructing a series of binary decisions. Starting from the root of the tree, the feature and threshold that maximize the separation between classes are selected. This is done recursively, selecting at each node the binary split rule that minimizes the class impurity. This algorithm has the advantage of being extremely interpretable, as it is easy to follow the sequence of rules that determined the classification of a given sample. Despite being a good algorithm for a baseline, its simplicity and reliance on hard rules makes it prone to overfit to its training dataset. 
One way to counter the lack of variance of these algorithms is to aggregate the predictions of multiple DTs. This is the strategy employed by Random Forests and Extreme Gradient Boost.

Random Forests~\cite{Breiman2001} are obtained by training several DTs in parallel, often using a random subset of all the available samples and also a subset of the existing features. Then, the predictions of the individual trees are aggregated through majority voting. This results in a more capable learning algorithm that is more robust to the noise in the data and less dependent on the specific constitution of the training dataset, often resulting in better performance.

Another way to ensemble multiple DTs is to use the gradient-boosted decision tree method~\cite{Friedman2001}, where DTs are added sequentially, with each new DT being fit on the error of the previous tree. XGBoost~\cite{Chen2016} is the most widely used implementation of this technique, which includes several improvements which often make it one of the best approaches for table-like data.

The Decision Trees and Random Forests were implemented, respectively, using the \texttt{scikit-learn} Python library~\cite{scikit-learn} \texttt{DecisionTreeClassifier} and \texttt{RandomForestClassifier}, while the XGBoost model was implemented using the \texttt{XGBClassifier} from the \texttt{xgboost} library~\cite{Chen2016}.

\subsubsection{Convolutional Neural Networks}
\label{subsec: resnet models}

For the image classification we employ models from the ResNet architecture family~\cite{He2016}. ResNets are convolutional neural networks (CNNs)~\cite{LeCun1989} which extract hierarchical features from images: early convolutional layers detect low-level patterns such as edges, while deeper layers capture higher-level, more abstract features. Convolutional kernels allow weight sharing across the multiple neurons, and their sparse connectivity reduces the number of parameters compared to fully connected networks, making CNNs efficient for high-dimensional inputs like spectrogram images. The defining feature of ResNets is the use of residual connections, which allow to train deeper models, as this stabilizes training and mitigates the vanishing gradients issue. We experimented with the ResNet18, ResNet34 and ResNet50 architectures, where the number (18, 34, 50) denotes the total number of convolutional layers in the respective architecture. 

The ResNets were implemented using the \texttt{pytorch-image-models} library~\cite{rw2019timm} and trained using \texttt{pytorch}~\cite{NEURIPS2019_9015}. For all models, the last layer was changed to have $C=7$ neurons (the number of glitch classes). The outputs of the final layer, known as logits, are passed through a softmax function $\sigma(\mathbf{z})_i = \frac{e^{z_i}}{\sum_{j=1}^{C} e^{z_j}}$, which converts them into scores that sum to one and can be interpreted as the probability of the input belonging to each class.

In the basic classification framework, a glitch is assigned to the class which receives the highest probability score, i.e. the \textit{argmax} of the softmax vector. 
As discussed later in section \ref{subsec: resnet results}, we define a minimum confidence threshold, below which a glitch is classified as \texttt{Low\_Confidence}.
This class is not intended as a mere residual category for model failures. Instead, it isolates events that cannot be reliably classified, quantifying the boundary of our current detector knowledge and serving as a guide for future detector characterization efforts and for continuous improvement of the classifier.

\section{Model training and selection}
\label{sec: model training}

Since our dataset is not balanced, we decided to not use accuracy as the main metric to compare the different models. Instead, we selected the F1 score, which corresponds to the harmonic mean of precision (the proportion of correct positive predictions among all predicted positives)
and recall (the proportion of actual positive examples that are correctly identified by the model),
\begin{equation}
    \label{eq: f1 score}
    \textrm{F1 score} = \frac{2} { \frac{1}{\textrm{precision}} + \frac{1}{\textrm{recall}} }  = \frac{\textrm{TP}} {\textrm{TP} + \frac{\textrm{FN} + \textrm{FP}}{2} },
\end{equation}
where TP, FP and FN denote, respectively, the total number of true positives, false positives and false negatives. This last equation yields one F1 score for each class. To obtain a single F1 score, we calculate the average accross all classes, which is known as the macro-averaged F1 score. For shortness, we will refer this simply as F1 score for the rest of the manuscript.

We used the training subset for training the models, and evaluated them in the validation dataset during the process of hyperparameter tuning, as detailed below. In the case of the XGBoost and ResNet models, a NVIDIA GeForce GTX 1050 Ti GPU was used to accelerate training.

\subsection{Decision Tree}

For the DT model, which serves as a baseline, we used the default parameters, except for the maximum tree depth, which was varied from 2 to 13. This parameter controls the complexity of the tree: deeper trees can capture more complex patterns in the data, but are also more prone to overfitting the training set. 

\begin{figure*}
    \centering
    \includegraphics[scale=1.3]{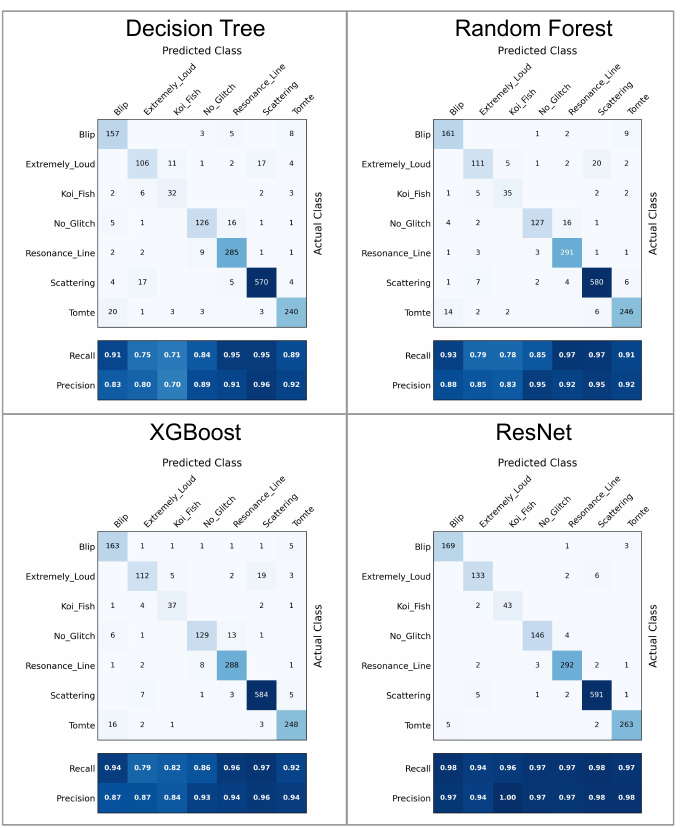}
    \caption{Confusion matrices for the validation dataset obtained with the best-performing configuration of each model (Decision Tree, Random Forest, XGBoost, and ResNet).
    The DT baseline shows several misclassifications, particularly for the minority \texttt{Koi\_Fish} class and between some class pairs. RF and XGBoost improve the classification performance, although some confusion between morphologically similar classes remains. The ResNet model provides the best overall results, with fewer misclassifications across all classes and a clear improvement for \texttt{Koi\_Fish} and previously problematic class pairs.}
    \label{fig: cms_valid}
\end{figure*}

The best DT, with a depth of 10, trained in 0.07 seconds and achieved a validation F1 score of {0.856} and an accuracy of {0.903}. The top-left confusion matrix in Figure~\ref{fig: cms_valid} shows that although this simple model provides a good baseline, with good recall and accuracy for some of the classes, it still struggles for the \texttt{Koi\_Fish} class. 
Furthermore, the model has trouble distinguishing some pairs of classes like \texttt{Extremely\_Loud} and \texttt{Scattering}, \texttt{Tomte} and \texttt{Blip}, or \texttt{No\_Glitch} and \texttt{Resonance\_Line}.

\subsection{Random Forest}

In the case of the RF models, hyperparameter tuning was performed using the \texttt{optuna} library~\cite{akiba2019optuna}, employing its default Tree-structured Parzen Estimator sampler, a probabilistic model-based approach for efficient search, with 100 trials. The allowed range and selected values for the hyperparameters are reported in Table~\ref{tab: rf_hp_tuning}. 
Parameter \texttt{n\_estimators} is the number of trees in the RF (more trees generally improve and stabilize predictions) and parameter \texttt{max\_depth} determines the depth of each individual DT, as deeper trees can capture higher complexity but are also more prone to overfitting. The \texttt{max\_features} and \texttt{min\_samples\_per\_leaf} parameters act as forms of regularization by limiting feature usage and preventing very small terminal nodes, while \texttt{bootstrap} introduces randomness in tree training to reduce variance and stabilize predictions. Finally, \texttt{class\_weight} adjusts the importance of classes to account for dataset imbalance.

\begin{table}[h!]
    \caption{List of Random Forest hyperparameters (first column), their allowed range (second column), and the tuned values for the glitch classification task.}
\begin{adjustbox}{width=0.5\textwidth}
    \centering
    \begin{tabular}{l|cc}
    \hline
    \textbf{hyperparameter} & \textbf{allowed values} & \textbf{selected value} \\
    \hline
    \texttt{n\_estimators} & [10, 200]  & 80 \\
    \texttt{max\_depth} & [3, 15]  & 14 \\
    \texttt{max\_features} & \texttt{"sqrt"}, \texttt{"log2"}, 0.5 or \texttt{None} & \texttt{"sqrt"} \\
    \texttt{min\_samples\_per\_leaf} & [1, 20] & 3\\ 
    \texttt{bootstrap} & \texttt{True} or \texttt{False} & \texttt{False}\\
    \texttt{class\_weight} & \texttt{None} or \texttt{"balanced"} & \texttt{None}\\
    \hline
    \end{tabular}
\end{adjustbox}
\label{tab: rf_hp_tuning}
\end{table}

The best RF model trained in 1.86 seconds and yielded a validation F1 score of {0.893} and an accuracy of {0.924}. As expected, these metrics are higher than the ones from the DT. 
The corresponding confusion matrix, displayed in the top-right panel of Figure~\ref{fig: cms_valid}, is cleaner than the previous one, with a noteworthy improvement for the \texttt{Koi\_Fish} class. Nevertheless, the confusion between some of the classes still persists. 

\subsection{XGBoost}

Similarly to the RF model, we performed hyperparameter tuning for the XGBoost model using \texttt{optuna} over 100 trials. The hyperparameter values used in the search and the chosen set are reported in Table~\ref{tab: xgb_hp_tuning}.
Here, \texttt{n\_estimators} defines the number of boosting rounds, with more iterations generally improving performance. The \texttt{learning\_rate} parameter controls how strongly each new tree contributes to the model; lower values make learning more gradual and help reduce overfitting. The \texttt{max\_depth} parameter determines the complexity of individual trees, as deeper trees capture more complex relationships but may overfit. The \texttt{min\_child\_weight} 
acts as a regularization term that discourages overly specific splits. Finally, \texttt{subsample} and \texttt{colsample\_bytree} control the fractions of training samples and features used for growing each tree, respectively, introducing randomness that improves generalization and reduces overfitting.

\begin{table}[h!]
\caption{List of XGBoost hyperparameters (first column), their allowed range (second column), and the tuned values for the glitch classification task.}
\begin{adjustbox}{width=0.5\textwidth}
    \centering
    \begin{tabular}{l|cc}
    \hline
    \textbf{hyperparameter} & \textbf{allowed values} & \textbf{selected value} \\
    \hline
    \texttt{n\_estimators} & [100, 1000]  & 450 \\
    \texttt{learning\_rate} & [0.01, 0.2]  & 0.0330 \\
    \texttt{max\_depth} & [1, 10] & 4 \\
    \texttt{min\_child\_weight} & [1, 7] & 6\\ 
    \texttt{subsample} & [0.5, 1.0] & 0.7811\\
    \texttt{colsample\_bytree} & [0.4, 1.0] & 0.8012\\
    \hline
    \end{tabular}
\end{adjustbox}
\label{tab: xgb_hp_tuning}
\end{table}

The best XGBoost model trained in 4.38 seconds and achieved a validation F1 score of {0.901} and an accuracy of {0.930}, which are higher than both previously reported models. 
The confusion matrix reported in the bottom left of Figure~\ref{fig: cms_valid} shows slight improvements for all classes, although some confusion between classes seen in the previous two confusion matrices persists. 
We performed a visual inspection of the spectrograms of the \texttt{Extremely\_Loud} samples predicted to be \texttt{Scattering}, \texttt{Blip} samples predicted to be \texttt{Tomte} and \texttt{No\_Glitch} samples predicted to be \texttt{Resonance\_Line}, and concluded the confusions are mostly due to limitations of the model rather than mislabeling. 
For the first pair of classes, we see that only 2 of the 19 glitches are indeed examples of \texttt{Scattering}. The remaining mostly fit the \texttt{Extremely\_Loud} class, although the majority is limited to very loud noise in the lower frequency regions. 
Regarding the second pair of classes, the confusion is expected as \texttt{Tomte} and \texttt{Blip} glitches are morphologically very similar, and could even be just two (overlapping) subsets of the same more general class. 
Finally, the confusion for the third pair of classes is due to spectrograms which are almost noiseless but show a small excess power at the region of the 500 Hz, although they do not look like the typical examples of \texttt{Resonance\_Line}.

\begin{figure}[!t]
    \centering
    \includegraphics[scale=0.44]{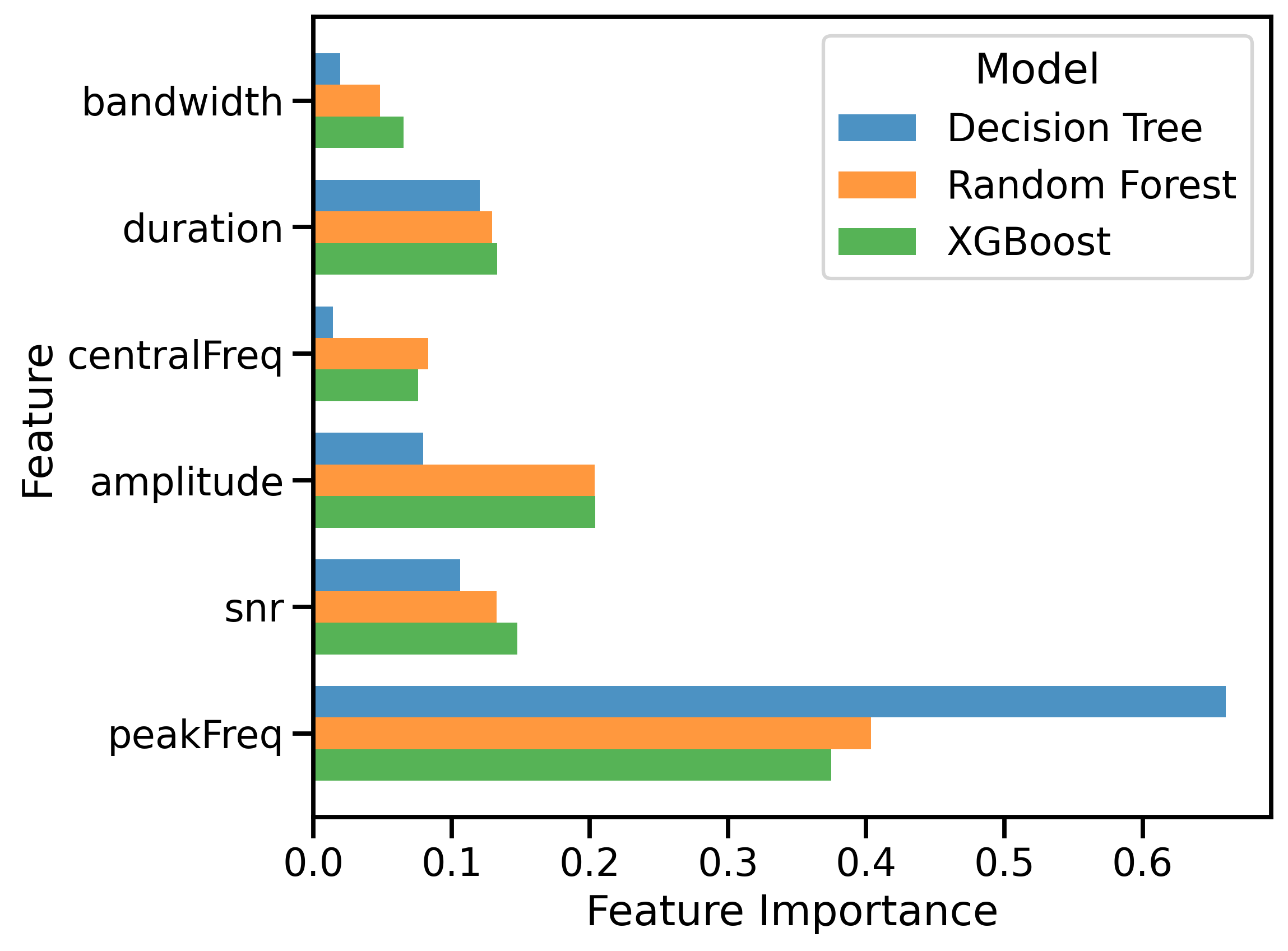}
    \caption{Feature importance of the \texttt{Omicron} parameters for each of the best tree-based models. While peak frequency is the primary predictor for all models, the ensemble methods (RF and XGBoost) exhibit a more distributed feature importance, reducing the dominance of peak frequency in favor of a more balanced contribution from SNR, amplitude, and duration.}
    \label{fig: fi_comparison}
\end{figure}

We made use of the interpretability of the tree-based models to investigate which features are most relevant for their glitch classification, as shown in  Figure~\ref{fig: fi_comparison}. 
The DT's predictions are mainly determined by the peak frequency, with smaller contributions from the duration, SNR and amplitude parameters. 
The RF is more balanced than the DT, as the importance of the peak frequency decreased 39\% and all the other features have increased their importance.
This is explained by the fact that at each splitting node of the RF, only a random subset of the features are considered, reducing the model's reliance on any single feature. In the case of the XGBoost, the feature importances indicate that the contribution of each feature is similar to the case of the RF, but slightly more balanced. Among the three models, XGBoost is the one that relies less on any single feature and achieves the highest accuracy. 

\subsection{ResNet}
\label{subsec: resnet results}

The ResNet models were trained using the cross-entropy loss, typical for multi-class classification problems:
\begin{equation}
    \label{eq: cross entropy loss}
    \mathcal{L} = - \frac{1}{N} \sum_{i=1}^{N} \sum_{k=1}^K {y}_k \log (\hat{p}_k).
\end{equation}
Here, $N$ is the total number of training samples, ${y}_k$ denotes the target probability of class $k$ (typically equal to 0 or 1), $K$ denotes the total number of classes (equal to the number of neurons in the output layer), and $\hat{p}_k$ is the predicted probability that a given example belongs to class $k$. 

Moreover, the networks were trained using the AdamW optimizer~\cite{Loshchilov2017-ej}, which combines the adaptive learning rate benefits of the Adam optimizer  with decoupled weight decay for better regularization, and a 1-cycle learning rate policy~\cite{Smith2018-no} that gradually increases the learning rate to a maximum value before annealing it down, which often accelerates convergence and improves generalization.

We also performed hyperparameter tuning for the ResNet model trained with the spectrograms, using \texttt{optuna} over 100 trials. Here, we optimized the \texttt{architecture}, to choose between the ResNet18, ResNet34 and ResNet50 versions. We also varied three main training hyperparameters, namely \texttt{n\_epochs}, the number of times the network processes the full training set, \texttt{batch\_size}, the number of samples the network processes in parallel at each training iteration, and \texttt{learning\_rate}, which controls the step size for updating the network's weights based on the error between predictions and ground truth. The values used for these hyperparameters and the chosen set are reported in Table~\ref{tab: resnet_hp_tuning}.

\begin{table}[h!]
\caption{List of ResNet hyperparameters (first column), their allowed range (second column), and the tuned values for the glitch classification task.}
\begin{adjustbox}{width=0.5\textwidth}
    \centering
    \begin{tabular}{l|cc}
    \hline
    \textbf{hyperparameter} & \textbf{allowed values} & \textbf{selected value} \\
    \hline
    \texttt{architecture} & ``ResNet18",``ResNet34" or ``ResNet50"  & ``ResNet34" \\
    \texttt{n\_epochs} & [7, 25]  & 17 \\
    \texttt{learning\_rate} & [$10^{-5}, 10^{-1}$] & $7.98 \times 10^{-3}$ \\
    \texttt{batch\_size} & [16, 128] & 32\\ 
    \hline
    \end{tabular}
\end{adjustbox}
\label{tab: resnet_hp_tuning}
\end{table}

\begin{figure}
    \centering
    \includegraphics[scale=0.44]{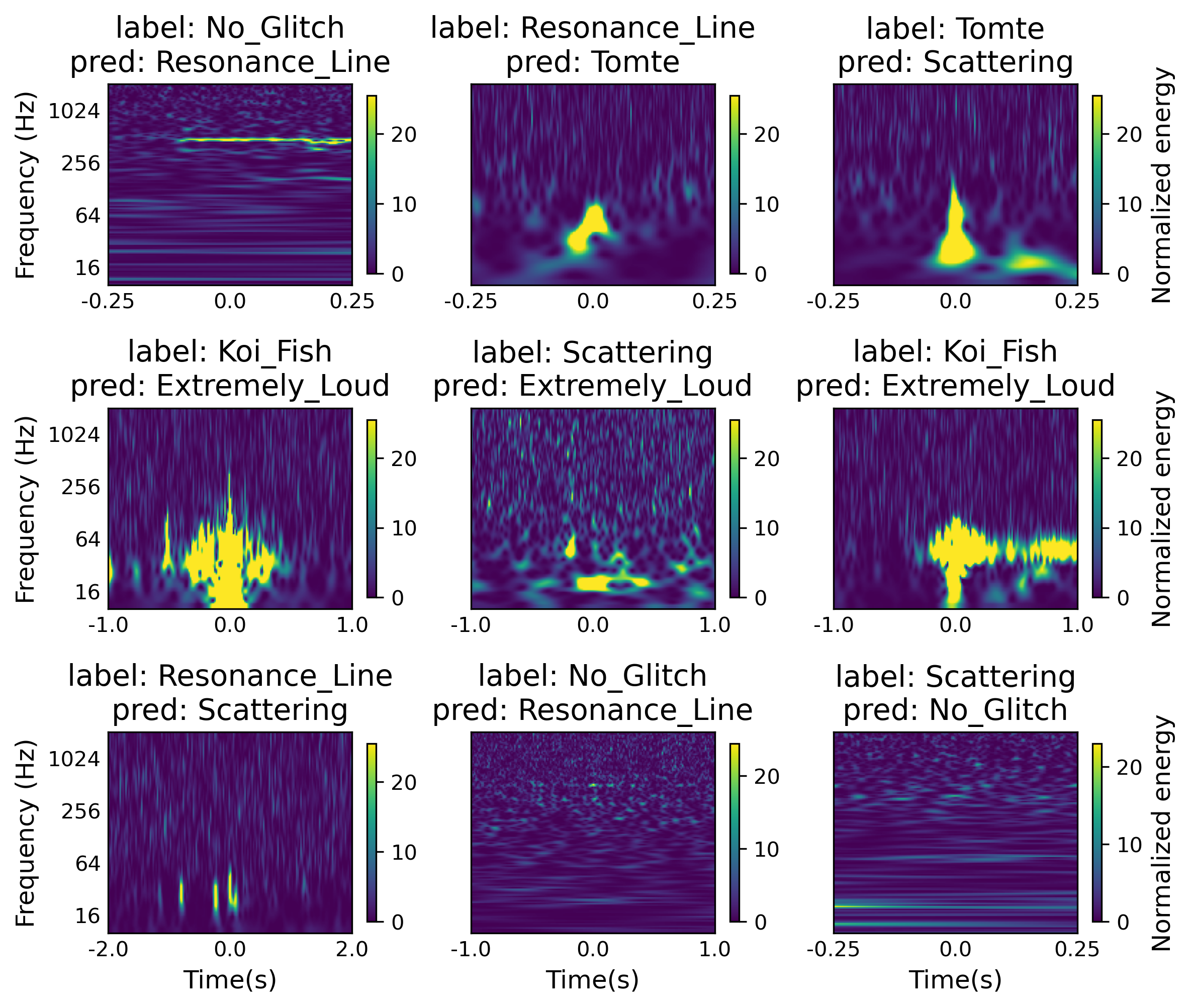}
    \caption{Glitches from the validation dataset where the ResNet predictions differed most from the ground truth. In most cases, the ground-truth labels appear incorrect, and the model predicts either the correct label or a more appropriate label.}
    \label{fig: resnet_top_losses_valid}
\end{figure}

The best ResNet achieved a {0.971} validation F1 score and a {0.975} accuracy, with a total training time of 940 seconds. 
The respective confusion matrix in the bottom right of Figure~\ref{fig: cms_valid} shows much less off-diagonal samples than the previous ones, as well as a notable improvement for the \texttt{Koi\_Fish} minority class. The three problematic pairs of classes, which previously had 19, 16, and 13 examples, now show only 6, 5, and 4 examples, respectively.
In addition, we show the samples where the ResNet's predictions were furthest from the ground truth in Figure~\ref{fig: resnet_top_losses_valid}. In most cases cases, the ground-truth label appears to be incorrect, and the model predicts either the correct label (e.g. top left) or a more appropriate label (e.g. top middle). For the event shown in the top right, we see that there is a \texttt{Tomte} at the middle of the window, which is immediately followed by a \texttt{Scattering}-like glitch, which could explain the model confusion. 
By observing the 42 glitches where the predicted label was different than the ground-truth label, it appears that the performance of the model, which is quite good, is not limited by the model itself, but by the quality of the labels.

Table~\ref{tab: valid_performance} summarizes the performance of the models in the validation dataset, as well as the time it took to train each model, $t_\textrm{train}$, and the time it takes to classify one glitch using the CPU, $t_\textrm{inf}$, which was averaged over 20 samples.
The model with the best F1 score is, by far, the ResNet. This more complex model, using a richer and more informative feature set, yielded a significant improvement in performance at the cost of longer training. However, inference remains fast, requiring on average only 0.04 seconds to process a single glitch even without GPU acceleration.

\begin{table}[t!]
\caption{Performance of the models in the validation dataset. The table reports the F1 score and accuracy, together with the training time ($t_\textrm{train}$) and the CPU inference time per glitch ($t_\textrm{inf}$). The ResNet achieves the highest performance by a large margin, at the cost of significantly longer training time, while maintaining a comparable inference time. }
    \centering
    \begin{tabular}{l|cccc}
    \hline
    \textbf{model} & \textbf{F1 score} & \textbf{accuracy} & \textbf{$t_\textrm{train}$ (s)} & \textbf{$t_\textrm{inf}$ (s)}\\
    \hline
    Decision Tree & 0.856 & 0.903 & 0.07 & $<0.01$\\
    Random Forest & 0.893  & 0.924 & 1.86 & $<0.01$ \\
    XGBoost & 0.901 & 0.930 & 4.38 & 0.01\\
    ResNet & \textbf{0.971} & 0.975 & 940 & 0.04\\ 
    \hline
    \end{tabular}
\label{tab: valid_performance}
\end{table}

We analyzed the distribution of the ResNet’s prediction confidences, i.e.~the maximum softmax probabilities on the validation set, to define a threshold for identifying low-confidence glitches. These low-confidence cases could correspond either to ambiguous glitches, whose class assignment is unclear among the known classes, or to novel glitches from previously unseen classes. The threshold for low-confidence glitches was set to the 2.5\% percentile of the validation scores, corresponding to 0.8543.

\section{Model evaluation}
\label{sec: model eval}

\begin{figure}
    \centering
    \includegraphics[scale=0.52]{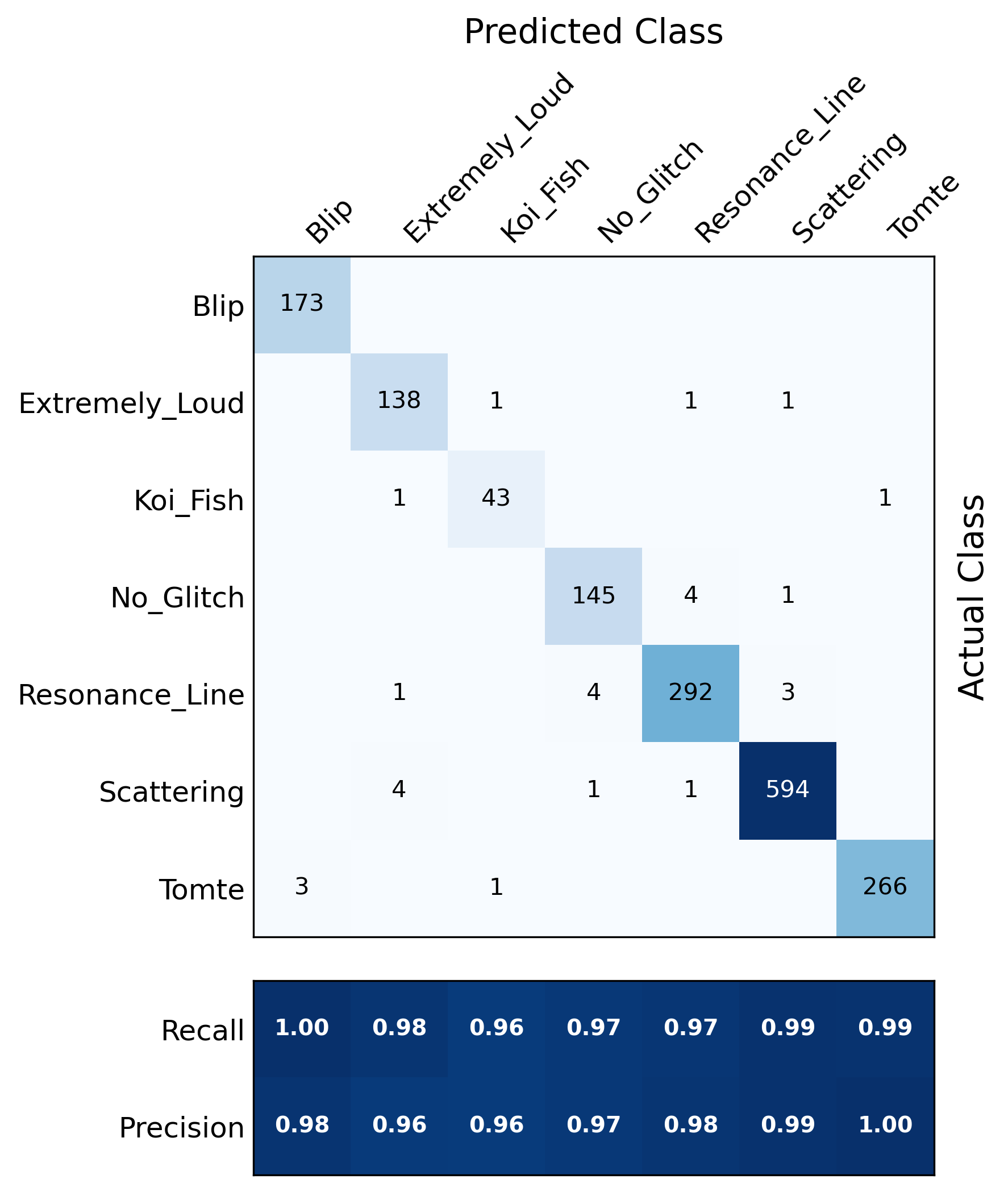}
    \caption{Confusion matrix obtained for the best ResNet model, for the held-out test dataset. The model shows precision and recall above 0.95 for every class.}
    \label{fig: cm_resnet_test}
\end{figure}

After choosing the best model using the validation dataset, we employed the test set, which had been kept apart until this time, to perform an unbiased estimation of the model's performance. 
We found that the model obtained an F1 score of {0.977} and an accuracy of {0.983}, which are actually slightly higher than in the validation dataset, confirming that it did not overfit. 
The confusion matrix of the predictions in Figure~\ref{fig: cm_resnet_test} shows that, from the 1679 glitches in the test dataset, only 28 are misclassified. Moreover, we see that the model performs very well even for the \texttt{Koi\_Fish} minority class, showing precision and recall above 0.95 for all classes. 
Once again, in the analysis of the few glitches where the predicted class did not match the ground truth, we found that the discrepancies are in most cases due to mislabeling rather than model error.

\begin{figure}[!htb]
    \centering
    \includegraphics[scale=0.44]{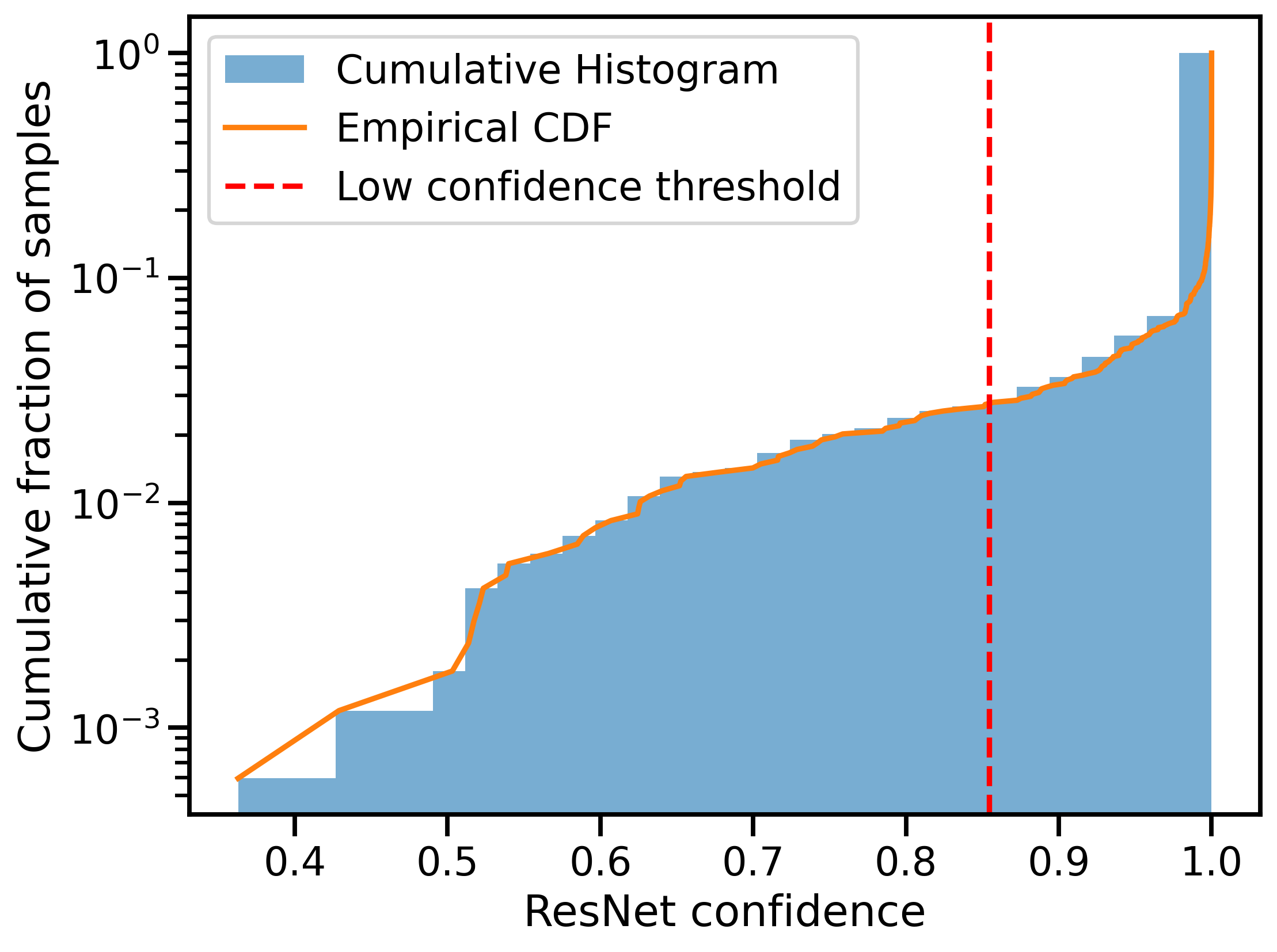}
    \caption{Distribution of the maximum softmax probabilities (confidence) of the ResNet model for the test dataset. The threshold used to identify low-confidence glitches is also shown. The model presents a clear skew to very high confidence levels.}
    \label{fig: resnet_test_confidence}
\end{figure}

\begin{figure}
    \centering
    \includegraphics[scale=0.5]{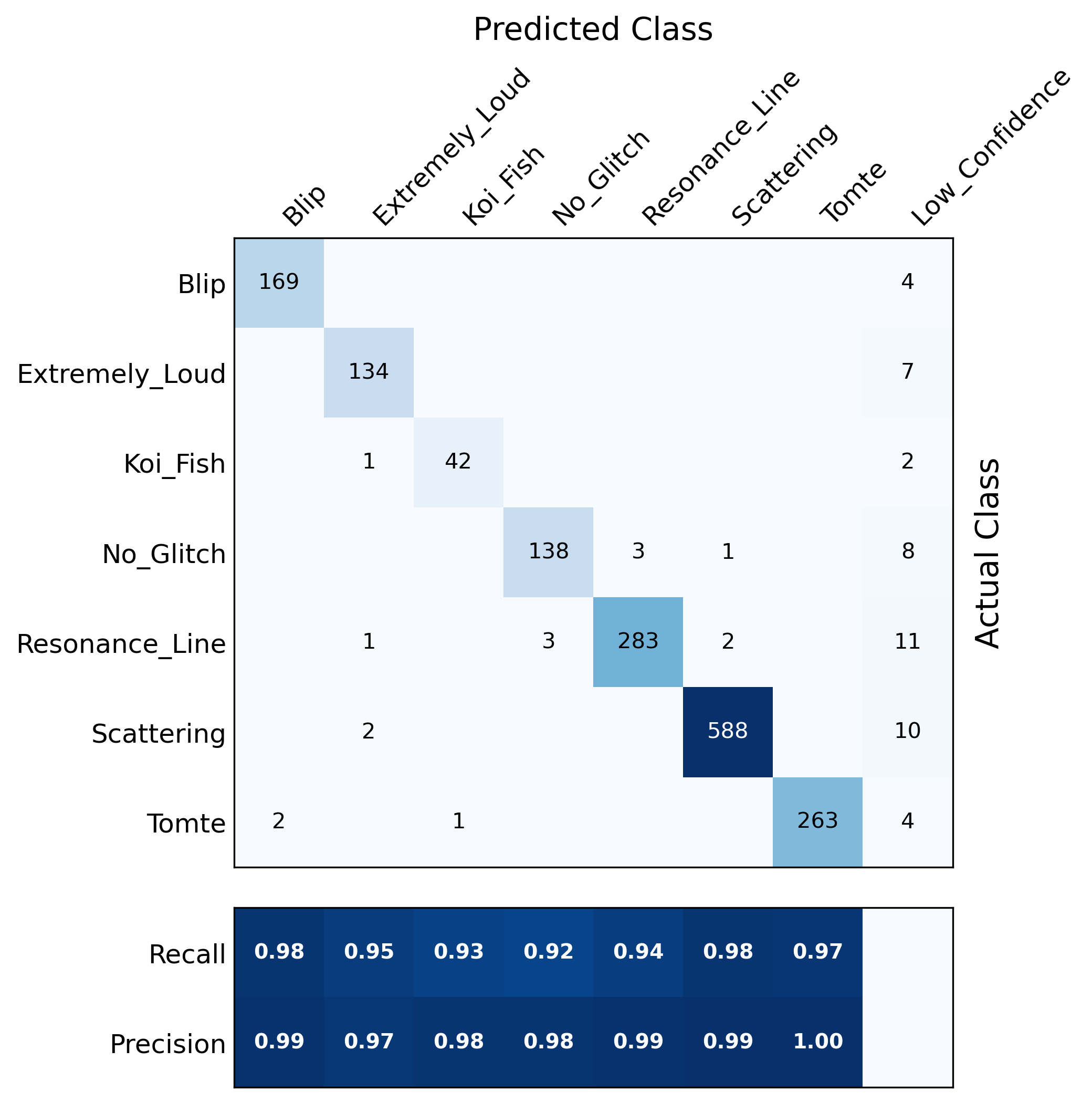}
    \caption{Confusion matrix obtained for the best ResNet model, for the test dataset, after applying the threshold to separate low-confidence glitches into a dedicated \texttt{Low\_Confidence} class. This increases the precision of the other classes while slightly reducing recall, and highlights the small fraction of glitches (2.7\%) where the model is less certain.}
    \label{fig: cm_resnet_test_lowconf}
\end{figure}

Moreover, we present the distribution of the confidence of our ResNet for the test glitches in Figure~\ref{fig: resnet_test_confidence}. The model, which is very accurate, is also highly confident in its predictions, with the median confidence at 0.999975 and the first quartile at 0.999650. 
Next, we applied the threshold defined using the validation dataset to single out the glitches where the model has lower confidence, as shown in Figure~\ref{fig: cm_resnet_test_lowconf}. 
This results in 46 glitches (2.7\% of the total) being labeled as \texttt{Low\_Confidence}. Of these 46 glitches, in around 70\% of the cases the low confidence is reasonable, either because they do not look like any class the model knows, or because they are near the boundary between classes. 
In the confusion matrix, we see that the inclusion of the \texttt{Low\_Confidence} class increases the precision for all the other classes, as it is more likely that glitches identified as some class do in fact belong to that class. 
Conversely, there is a small decrease in recall, as some of the glitches that were correctly labeled as belonging to a given class now are classified as \texttt{Low\_Confidence}. 
Thus, this thresholding results in more confident classifications, while setting apart a small percentage of the glitches for further analysis.

\begin{figure}
    \centering
    \includegraphics[scale=0.64]{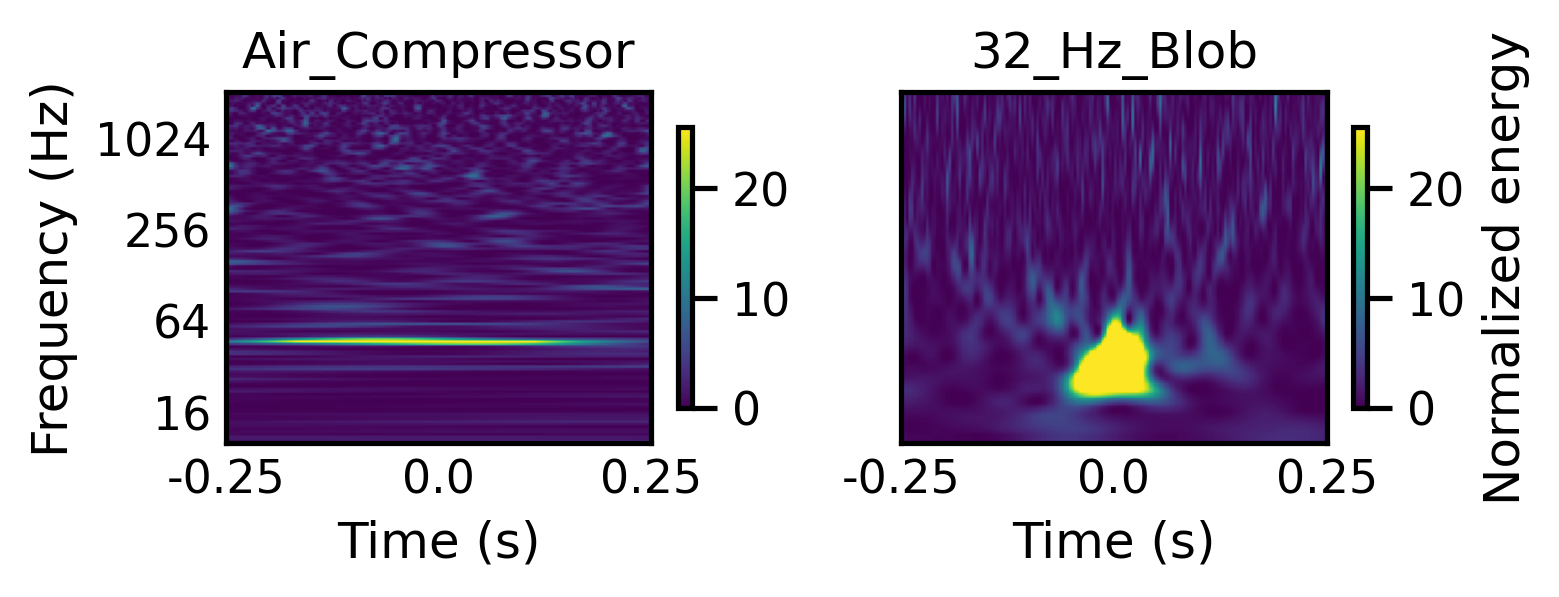}
    \caption{Examples of glitches from two out-of-domain classes used to test the generalization of the ResNet model. The first class, \texttt{Air\_Compressor}, shows short thick lines around 50 Hz, while the second class, \texttt{32\_Hz\_Blobs}, contains mostly round shapes centered at 32 Hz.}
    \label{fig: out_domain_classes}
\end{figure}

Finally, to test the generalization of the model to unseen glitch classes, we collected two sets of 100 morphologically related glitches from the initial O3b glitch dataset. 
The first set consists of the 100 most confident examples of Gravity Spy's \texttt{Air\_Compressor} class. Originally, this class included LIGO Hanford glitches showing a short, thick line centered at 50 Hz, caused by the switching on and off of the air compressor motors, and was removed during O1 after improving the vibration isolation~\cite{Bahaadini2018}. 
In more recent runs, Gravity Spy attributes this label to glitches that are morphologically similar, showing a line around 50 Hz.  
The second set contains glitches that show an irregular but mostly round shape centered at 32 Hz, which we call \texttt{32\_Hz\_Blobs}. These were collected by visual inspection of the most confident glitches labeled as \texttt{None\_of\_the\_Above}. Figure~\ref{fig: out_domain_classes} shows a representative example from each set.
In an ideal scenario, because these two noise morphologies are out-of-distribution for our supervised model, we expect the model to yield maximum probability scores lower than our confidence threshold for all such glitches. Consequently, they would be categorized as \texttt{Low\_Confidence} and would not contaminate the known classes.

\begin{figure}
    \centering
    \includegraphics[scale=0.42]{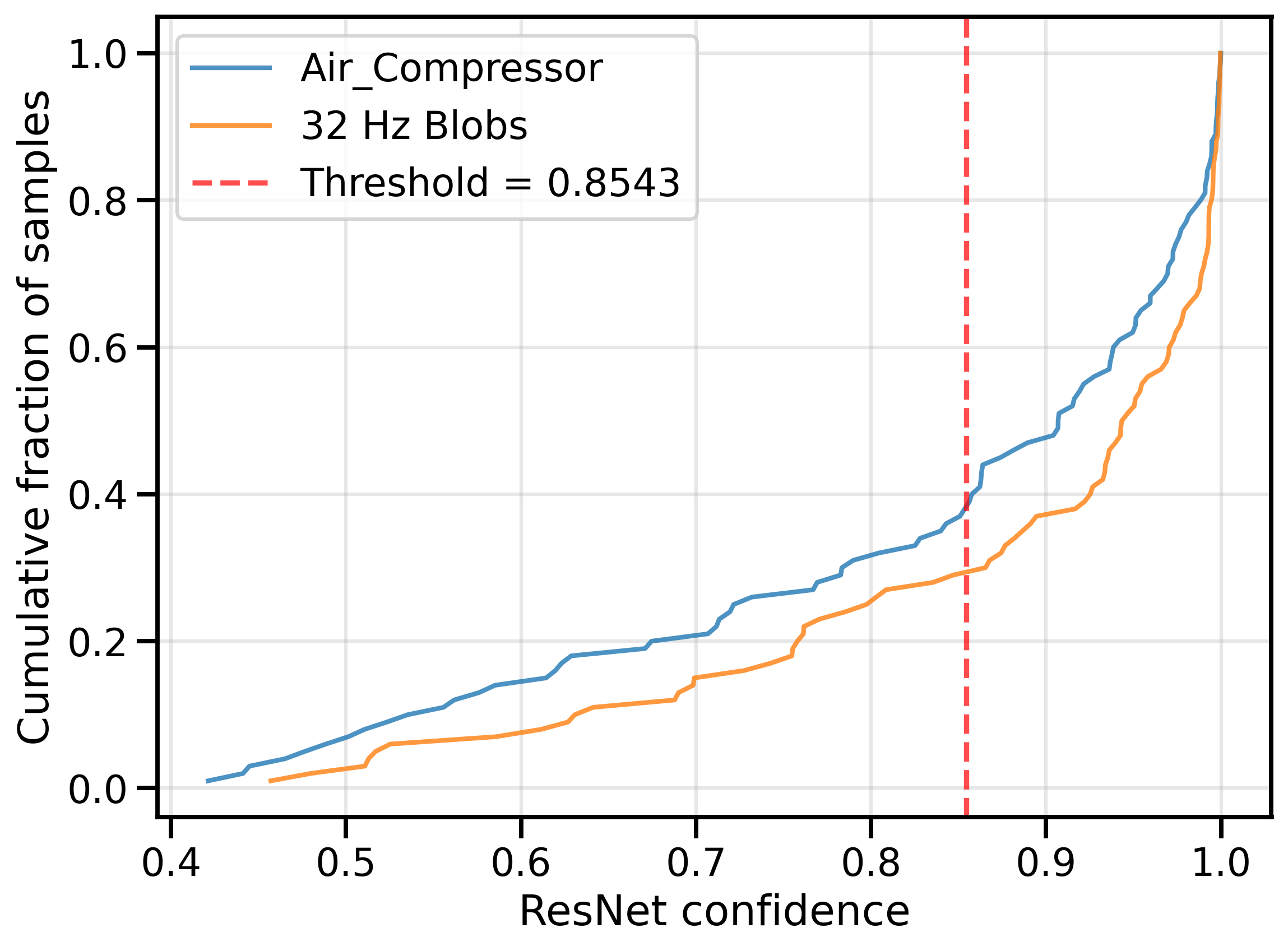}
    \caption{Distribution of the maximum softmax probabilities (confidence) for the ResNet model, for the two out-of-domain classes. Glitches with confidence below the threshold are flagged as \texttt{Low\_Confidence}, corresponding to 38\% of \texttt{Air\_Compressor} glitches and 29\% of \texttt{32\_Hz\_Blobs}.}
    \label{fig: out_domain_confidence}
\end{figure}

We obtained the model predictions and softmax probabilities for the two sets of glitches in order to evaluate the degree of confidence the model would have in these out-of-domain glitches, and which labels it would predict. 
We show the distribution of the model confidences in Figure~\ref{fig: out_domain_confidence}. We find that 38\% of \texttt{Air\_Compressor} glitches have a confidence below the threshold, while in the case of the \texttt{32\_Hz\_Blobs}, 29\% are below the confidence threshold. These percentages correspond to the portion of glitches flagged as \texttt{Low\_Confidence}, meaning that they would require further attention.

\begin{figure}[!htb]
    \centering
    \includegraphics[scale=0.54]{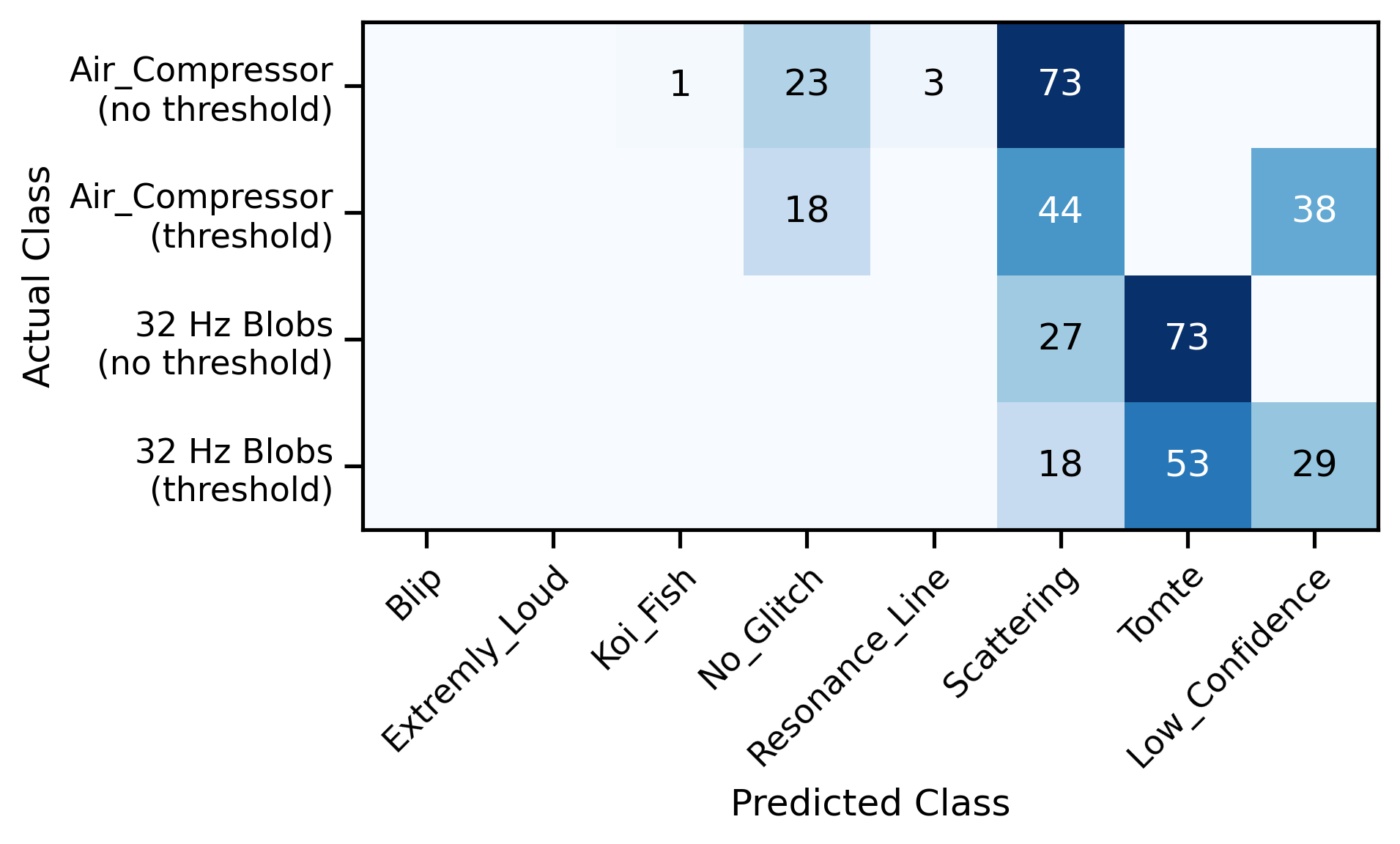}
    \caption{Confusion matrix for the out-of-domain glitches, before and after applying the threshold on the model confidence. After thresholding, \texttt{Air\_Compressor} glitches are mostly classified as \texttt{Scattering} or \texttt{No\_Glitch}, while \texttt{32\_Hz\_Blobs} are labeled as \texttt{Tomte} or \texttt{Scattering}.}
    \label{fig: out_domain_cm}
\end{figure}

We also show in Figure~\ref{fig: out_domain_cm} the confusion matrices for each of the sets, before and after applying the confidence thresholding. We observe that the \texttt{Air\_Compressor} glitches were mostly identified as \texttt{Scattering} and \texttt{No\_Glitch}. We observed that cases showing an horizontal line with around half second of duration were erroneously classified as \texttt{Scattering}, while cases where we see a longer-duration (up to 4 seconds) spotty line around 50 Hz were classified as \texttt{No\_Glitch}. In the case of the \texttt{32\_Hz\_Blobs}, they were mostly classified as \texttt{Tomte}, when the shape was more circular, or \texttt{Scattering}, when the glitch was more elongated. For both sets of glitches, we could not identify any morphological feature that consistently explained the differences in confidence between pairs of similar glitches.
These results highlight a limitation of purely supervised models when applied to out-of-distribution glitches, as instead of producing low softmax probabilities for unfamiliar morphologies, the network often assigns high confidence to known classes with loose similarities.

\section{Pipeline implementation in Virgo}
\label{sec: pipeline}

\texttt{VIGILant} runs daily on a computing node at the Computing Center of the European Gravitational Observatory (site of the Virgo detector in Cascina), generating glitch classifications for the previous day and updating the glitch dashboard. The process takes approximately 3 seconds per glitch, mostly due to the spectrogram creation step.

In essence, the pipeline performs the following steps:

\begin{enumerate}
    \item fetch the \texttt{Omicron} unclustered triggers from the previous day;
    \item cluster the triggers;
    \item filter out clusters below an SNR threshold of $7.5$
    \item generate spectrograms;
    \item use the trained ResNet to get the model predictions;
    \item update the glitch dashboard.
\end{enumerate}

Initially, \texttt{VIGILant} retrieves all \texttt{Omicron} unclustered triggers from the previous day, with a 10-second margin to avoid splitting glitches that occur around the change of the calendar day. Furthermore, only triggers produced while the Virgo interferometer was in  observing mode, i.e.~officially collecting data for astrophysical analysis, are considered.

Then, these triggers are clustered in time and frequency using a two-pass approach. First, temporal clustering groups events that fall within a 0.1 s window. Second, each temporal cluster is subdivided in frequency if the triggers’ frequency bands are separated by more than 10 Hz. Finally, the algorithm aggregates each cluster into a single representative event, adopting the start and end bounds of the group while inheriting the GPS time, peak frequency, SNR and amplitude from the trigger with the highest SNR within that cluster. 
Following the Gravity Spy criterion, we retain only merged clusters with a peak SNR $\ge 7.5$ for spectrogram generation and classification, to discard low-significance noise fluctuations.

Following that, three spectrograms (corresponding to the three time windows used so far) are generated for each glitch, following the procedure discussed in subsection~\ref{subsection: spectrograms}.  These are then combined using the ``encoded134" approach and processed by the trained ResNet, which outputs the probabilities for each glitch class. Using these probabilities, the glitch is labeled as the class with maximum probability, if it is higher than the previously defined threshold, or as a \texttt{Low\_Confidence} glitch otherwise. 
These predictions, along with the \texttt{Omicron} parameters, are saved into a CSV file  to allow their use in further investigations.

\begin{figure*}[!htbp]
    \centering
    \includegraphics[scale=0.72]{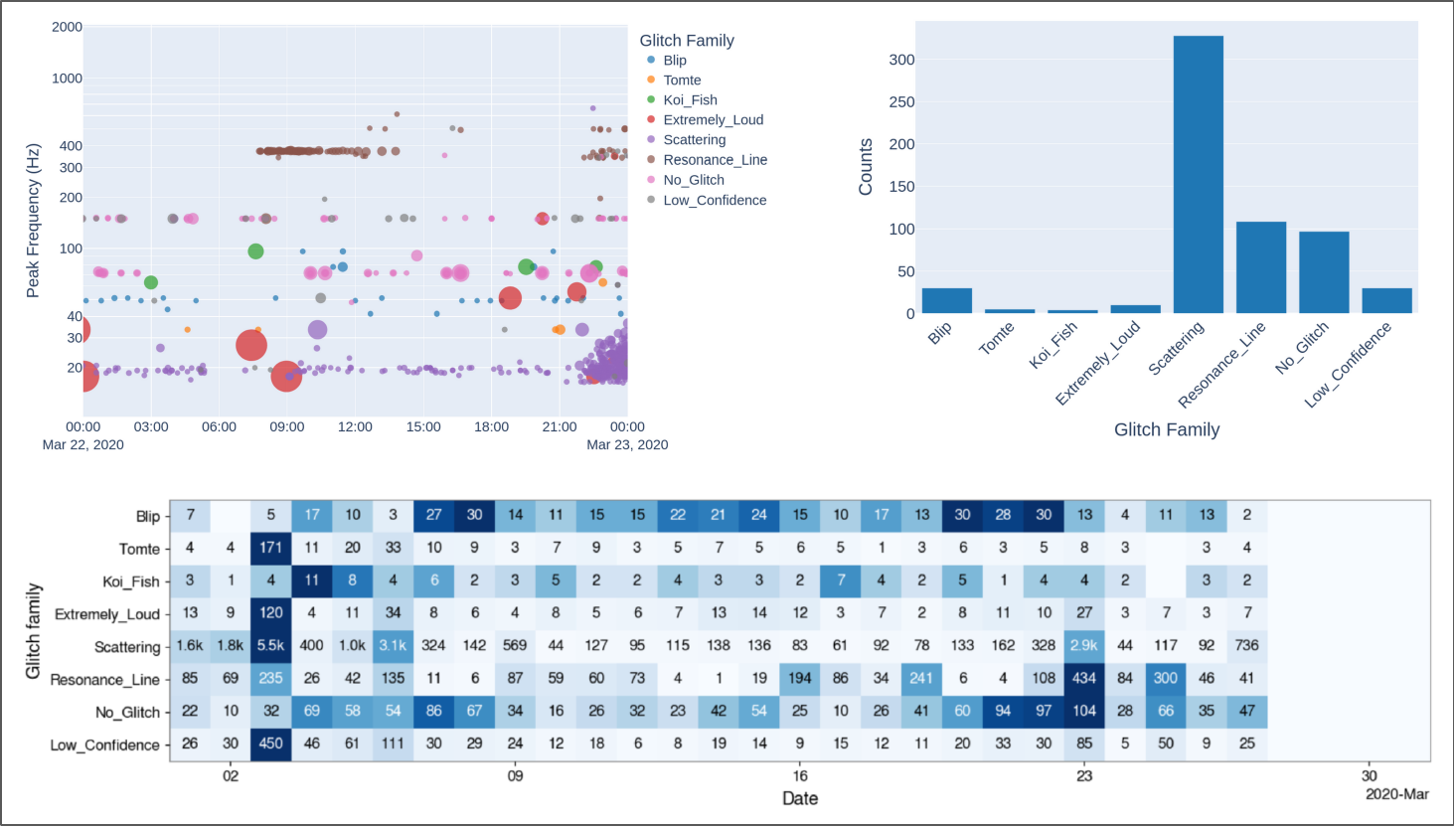}
    \caption{Example plots from the glitch dashboard for March 22, 2020. The top left panel shows the glitchgram, with the glitch time in the $x$-axis and the peak frequency in the $y$-axis;  marker size encodes SNR, and marker color indicates the glitch family. The top right panel shows the daily glitch count bar plot, reporting the number of glitches per family on that day. Finally, the bottom panel shows the monthly glitch heatmap for March 2020, summarizing daily glitch counts for each family. This highlights unusual behavior, such as March 16, which had relatively few glitches overall but a disproportionately high number of \texttt{Resonance\_Line} glitches.
    An interactive version of the dashboard is accessible at \url{https://tttiago.github.io/glitchgram/dashboard_demo.html}.}
    \label{fig: glitch dashboard}
\end{figure*}

Finally, the glitch predictions are aggregated in a monthly dashboard, which presents three plots. The first one is an enhanced glitchgram, shown in the top-left panel of Figure~\ref{fig: glitch dashboard}. This plot, like the classic glitchgram, also displays the glitches in the time and frequency dimensions and encodes their SNR in the marker size\footnote{The classic glitchgram uses a color-scale to encode the SNR.}. Notably, it also displays the label of the glitch using the color of the marker. Moreover, the use of the \texttt{plotly} library allows this glitchgram to be interactive, giving the possibility of selecting only some glitch classes, shortening or extending the limits in the time and/or frequency axes, and hovering each marker to get more complete information about each specific glitch.
The second plot, shown in the top-right panel of Figure~\ref{fig: glitch dashboard}, is a bar plot which shows the counts of each glitch family for a given calendar day, providing a quick overview of the detector behavior over that period of time. Finally, a heatmap as the one shown in the bottom panel of Figure~\ref{fig: glitch dashboard} summarizes the information across a whole month, displaying daily counts for each glitch family and using a color scale to facilitate the identification of outliers. This plot enables monitoring of the detector’s glitch activity over time and facilitates the identification of changes in the rates of individual glitch classes.

The tool, as well as instructions for access to its outputs, can be found at \url{https://git.ligo.org/tiago.fernandes/vigilant}.

\section{Conclusions}
\label{sec: conclusion}

We have introduced \texttt{VIGILant}, an automatic pipeline for low-latency classification and visualization of glitches in the Virgo detector. This tool combines a Machine Learning model that classifies glitches with a graphical interface to help the visualization of the glitch distribution in time, frequency, SNR and glitch family. This pipeline has been running daily since the O4c observation run of the LVK Collaboration, providing an important detector characterization product to the Virgo collaboration.

Two types of techniques have been evaluated for the classification: i) tree-based models using the \texttt{Omicron} parameters, and ii) convolutional neural networks performing image classification on the glitch spectrograms. It has been found that Random Forest and XGBoost were quick to train and reasonably accurate, especially when considering the limited input information. CNNs, however, achieve much higher accuracy. This is essentially limited only by the quality of the labels in the training dataset, at the cost of a longer training time, but still maintaining an inference time on the order of tens of milliseconds. 

For the creation of the Virgo O3b glitch dataset, we combined the Gravity Spy model's labels with the labels from the citizen science and visual inspection of the glitches. This resulted in a dataset with more accurate labels that better reflects the glitch population in Virgo. Building the dataset highlighted one limitation of supervised deep learning methods such as CNNs, which require large and well-labeled datasets for training.
While manual expert verification can mitigate label noise, checking every glitch in a dataset with the size typically needed for training these models is infeasible due to time and person-power constraints.
Furthermore, detector evolution across observing runs inevitably introduces morphological shifts and novel glitch classes. Although core glitch families represent fundamental signal morphologies that remain broadly effective across runs, periodic retraining can help the model better adapt to subtle morphological shifts.
In this regard, the exploration of methods less dependent on manual labor, such as the use of unsupervised clustering algorithms, could be useful for the creation of labeled datasets for supervised classification.

We have observed that the supervised model struggles with out-of-domain glitches and produces overly confident classifications. While confidence thresholding still catches 30--40\% of out-of-domain glitches as \texttt{Low\_Confidence}, the model often assigns high confidence to glitches from novel families. This underscores the need for dedicated unsupervised anomaly-detection methods to flag new glitch classes. 
Another area of possible improvement is the use of glitch representations less dependent on the choice of Q. Instead of using a single Q-transform to represent each glitch, a promising possibility is to stack the multiple Q-tiles produced by \texttt{Omicron}. This could give a less biased view of each glitch and improve its classification.

Looking forward, \texttt{VIGILant} is intended to go beyond glitch classification by enabling the identification of new glitch classes and the reconstruction and subtraction of glitch signals. New glitch classes could be identified by clustering low-confidence samples (flagged by our current algorithm as \texttt{Low\_Confidence}) or by incorporating unsupervised anomaly detection in order to propose new classes once enough examples are found, using for instance a method similar to~\cite{Laguarta2023}. These new classes could then be periodically added to the supervised classifier’s training set, keeping it up-to-date with the ever evolving glitch population.
For glitch reconstruction and subtraction, we plan to make use of total variation methods and Sparse Dictionary Learning, following the work laid out in~\cite{Torres-Forne2020,clawdia}. These extensions, left for future work, would provide a complete pipeline focused on the identification, analysis, and mitigation of glitches in the Virgo detector.

\section*{Data availability statement}

The data that support the findings of this article are openly available \cite{vigilant-analysis-repo}.

\begin{acknowledgments}

We are thankful to Osvaldo Freitas, Solange Nunes, Nino Villanueva, Anastasios Theodoropoulos, and José Martin-Guerrero for their helpful comments and suggestions during our weekly group meetings.
We also express our gratitude to Guillem Fernandéz, Melissa López and Isabel Cordero-Carrión for insightful discussions regarding the theme of this work.
We also thank Marco Cavaglià for the internal LIGO-Virgo-KAGRA review.

TF is supported by FCT - Fundação para a Ciência e Tecnologia, I.P. through doctoral scholarship 2023.03753.BD (\url{https://doi.org/10.54499/2023.03753.BD}).
TF and AO acknowledge financial support by CF-UM-UP through the Strategic Fundings UIDB/04650/2020, UIDP/04650/2020, UID/PRR/04650/2025 and UID/04650/2025 of FCT.
JAF, ATF, and TF are supported by the Spanish Agencia Estatal de Investigaci\'on (grant PID2024-159689NB-C21) funded by MICIU/AEI/10.13039/501100011033 and by FEDER/EU. JAF and ATF are also supported by the Generalitat Valenciana (Prometeo grant CIPROM/2022/49), and by  the  European Horizon  Europe  staff  exchange  (SE)  programme HORIZON-MSCA-2021-SE-01 (grant NewFunFiCO-101086251). 

We acknowledge the use of open source \texttt{python} libraries including \texttt{numpy}~\cite{harris2020array}, \texttt{pandas}~\cite{mckinney-proc-scipy-2010}, \texttt{matplotlib}~\cite{Hunter:2007}, \texttt{plotly}~\cite{plotly}, \texttt{scikit-learn}~\cite{scikit-learn}, \texttt{xgboost}~\cite{Chen2016} \texttt{umap-learn}~\cite{McInnes2018-xo}, \texttt{gwpy}~\cite{gwpy}, \texttt{pytorch}~\cite{NEURIPS2019_9015}, \texttt{torchvision}~\cite{torchvision2016}, \texttt{pytorch-image-models}~\cite{rw2019timm}, \texttt{pillow}~\cite{clark2015pillow} and \texttt{optuna}~\cite{akiba2019optuna}.

This material is based upon work supported by NSF’s LIGO Laboratory which is a major facility fully funded by the National Science Foundation, as well as the Science and Technology Facilities Council (STFC) of the United Kingdom, the Max-Planck-Society (MPS), and the State of Niedersachsen/Germany for support of the construction of Advanced LIGO and construction and operation of the GEO600 detector. Additional support for Advanced LIGO was provided by the Australian Research Council. Virgo is funded, through the European Gravitational Observatory (EGO), by the French Centre National de Recherche Scientifique (CNRS), the Italian Istituto Nazionale di Fisica Nucleare (INFN) and the Dutch Nikhef, with contributions by institutions from Belgium, Germany, Greece, Hungary, Ireland, Japan, Monaco, Poland, Portugal, Spain. KAGRA is supported by Ministry of Education, Culture, Sports, Science and Technology (MEXT), Japan Society for the Promotion of Science (JSPS) in Japan; National Research Foundation (NRF) and Ministry of Science and ICT (MSIT) in Korea; Academia Sinica (AS) and National Science and Technology Council (NSTC) in Taiwan. 
The authors gratefully acknowledge the support of the NSF, STFC, INFN, CNRS and Nikhef for provision of computational resources.

\end{acknowledgments}

\bibliographystyle{apsrev4-1}
\bibliography{references}

\end{document}